\documentclass[aps,prd,superscriptaddress,twocolumn,nofootinbib]{revtex4-2}
\usepackage[normalem]{ulem}
\usepackage{natbib}
\usepackage{graphicx}
\usepackage{amsfonts}
\usepackage{amsmath,amssymb} % \colleneqq for definition symbol
\usepackage{bm}
\usepackage{xcolor}
\usepackage{hyperref}
\usepackage[caption=false]{subfig}
\usepackage{tabularx}
\usepackage{mathtools}
\usepackage{xfrac} % for slanted fractions

\newcommand{\En}{\mathcal{E}}
\newcommand{\Lz}{{\mathcal{L}_z}}
\newcommand{\Q}{\mathcal{Q}}
\newcommand{\K}{\mathcal{K}}
\newcommand{\sn}{\mathrm{sn}}
\newcommand{\am}{\mathrm{am}}
\renewcommand{\sp}{\epsilon} %%Small Parameter = mass ratio
\newcommand{\mr}{\epsilon} %% Mass Ratio
\renewcommand{\r}{r_p}
\newcommand{\z}{z_p} 
\renewcommand{\th}{{\theta_p}} 
\newcommand{\ph}{{\phi_p}}
\renewcommand{\t}{{t_p}}
\newcommand{\pos}[1]{#1_p}
\newcommand{\nitr}{{\Tilde{r}_p}}
\newcommand{\tnitr}{{r_\varphi}}
\newcommand{\tter}[1]{{r^{(#1)}_\varphi}}

\newcommand{\slowt}{{\mathcal{T}}}

\newcommand{\mycomment}[1]{} %Command to effectively comment out blocks of code in TeX

\newcommand{\avg}[1]{\left\langle #1 \right\rangle}
\newcommand{\bavg}[1]{\big\langle #1 \big\rangle}
\newcommand{\osc}[1]{\Breve{#1}}
\newcommand{\nit}[1]{\Tilde{#1}}
\newcommand{\PD}[2]{\frac{\partial #1}{\partial #2}}
\newcommand{\HOT}[1]{\mathcal{O}( \sp ^{ #1 } )} 

\begin{document}

%\preprint{}

\title{Self-forced inspirals  with spin-orbit precession}


\newcommand{\NBIA}{Niels Bohr International Academy, Niels Bohr Institute, Blegdamsvej 17, 2100 Copenhagen, Denmark}
\newcommand{\AEI}{Max Planck Institute for Gravitational Physics (Albert Einstein Institute), Am M\"uhlenberg 1, Potsdam 14476, Germany}
\newcommand{\UCD}{School of Mathematics and Statistics, University College Dublin, Belfield, Dublin 4}

\author{Philip Lynch}
%\email[Email: ]{philip.lynch@ucdconnect.ie}
%\thanks{}
\affiliation{\AEI}
\affiliation{\UCD}
\author{Maarten van de Meent}
\affiliation{\AEI}
\affiliation{\NBIA}
\author{Niels Warburton}
\affiliation{\UCD}

\date{\today}

\begin{abstract}

We develop the first model for extreme mass-ratio inspirals (EMRIs) with misaligned angular momentum and primary spin, and zero eccentricity ---also known as quasi-spherical inspirals--- evolving under the influence of the first-order in mass ratio gravitational self-force. 
The forcing terms are provided by an efficient spectral interpolation of the first-order gravitational self-force in the outgoing radiation gauge. 
In order to speed up the calculation of the inspiral we apply a near-identity (averaging) transformation to eliminate all dependence of the orbital phases from the equations of motion while maintaining all secular effects of the first-order gravitational self-force at post-adiabatic order.
The resulting solutions are defined with respect to `Mino time' so we perform a second averaging transformation so the inspiral is parametrized in terms of Boyer-Lindquist time, which is more convenient for LISA data analysis.
We also perform a similar analysis using the two-timescale expansion and find that using either approach yields self-forced inspirals that can be evolved to sub-radian accuracy in less than a second.
The dominant contribution to the inspiral phase comes from the adiabatic contributions and so we further refine our self-force model using information from gravitational wave flux calculations.
The significant dephasing we observe between the lower and higher accuracy models highlights the importance of accurately capturing adiabatic contributions to the phase evolution.
\end{abstract}

%\keywords{}

\maketitle


\section{Introduction \label{section:Intro}}
With the advent of spaced-based gravitational wave~(GW) detectors such as Laser Interferometer Space Antenna (LISA) \cite{Baker:2019nia}, TianQin \cite{TainQin}, and Taiji \cite{Teiji}  in the 2030s, comes the need to model sources in the millihertz frequency range. 
A prime source for these future detectors are extreme mass ratio inspirals (EMRIs).
These systems consist of a massive black hole (MBH) primary with mass $M$ and a stellar-mass compact object secondary with mass $\mu$ (e.g., a steller-mass black hole or neutron star).
These space based detectors will be to be sensitive to EMRIs with mass ratios $\mr = \mu/M \sim  10^{-4} - 10^{-7}$. 
Gravitational waves from these systems are expected to stay in the LISA band for months to years with the secondary completing $\sim 1/\mr = 10^4 - 10^7$ orbits spiralling around the primary before merging under the influence gravitational radiation reaction.
Accurately modelling these systems through such a large number of orbital cycles to sub-radian accuracy with methods efficient enough for parameter estimation remains an open challenge, but one that will grant precise parameter estimation for MBHs along with some of the most sensitive probes for new physics beyond general relativity~\cite{LISAConsortiumWaveformWorkingGroup:2023arg,Berry2019,Gair:2012nm}.

In this work, we look at the subclass of these systems which have zero eccentricity and misaligned orbital angular momentum and primary spin. In the strong field regime this will cause the orbital plane to rapidly precess around the primary's spin, ergodically filling (part of) a sphere, if we were to ignore orbital evolution due the emission of gravitational waves. Consequently, such systems are also known as quasi-spherical inspirals.
 One probable EMRI formation channel is the tidal disruption of a comparable mass black hole binary by a nearby MBH which will result in a circular and arbitrarily inclined inspiral by the time it enters the LISA band \cite{Babak:2014kqa,Hopman_2009,ColemanMiller:2005rm}.
While the primary EMRI  formation channel of direct capture is expected to produce LISA band inspirals with eccentricity and inclination, radiation reaction will cause an EMRI to circularize (with the exception of right before the last stable orbit) \cite{Glampedakis2002} and cause its inclination to slowly increase \cite{Ryan:1995xi}.
Thus an EMRI with any  amount of inclination but low eccentricity is likely to evolve into a quasi-spherical inspiral before plunge.

The leading order behaviour of quasi-spherical EMRIs have been modelled by calculating the asymptotic fluxes at infinity and through the horizon of the primary and relating these to the averaged rate of change of energy, angular momentum, and Carter constant of the system \cite{Hughes2000,Hughes2001,Hughes2021}. 
This same approach has recently been generalized to inspirals with inclination and eccentricity~\cite{Hughes2021}.
While such inspiral calculations are numerically efficient and maybe accurate enough for detection of loud EMRI  signals, the resulting adiabatic (0PA) inspirals do not reach the sub-radian accuracy requirement needed to reach all of LISA's science goals~\cite{Amaro-Seoane:2011fgy}.

For that, one must understand the sub-leading order in mass ratio behaviour and compute post-adiabatic (1PA)  inspirals.
This will require knowledge of the local force on the secondary due to the presence of its own gravitational field, i.e., its gravitational self-force (GSF).
This is computed by expanding the metric of the binary as the Kerr metric of the primary plus perturbations which are of expressed as a power series of the small mass ratio $\mr$~\cite{Barack:2018yvs, Poisson:2011nh,Pound2021}.
Using a two-timescale analysis, it has been determined that to reach sub-radian accuracy across the inspiral, one will need to know all first-order in the mass ratio effects of the GSF as well as the orbit averaged dissipative contribution from the second-order in the mass ratio GSF~\cite{Hinderer2008}. 
While second-order results for Schwarzschild spacetime are now emerging~\cite{Pound2019,Warburton2021,Durkan:2022fvm,Wardell:2021fyy,Burke:2023lno}, calculations for inspirals in Kerr spacetime are likely still a few years away.
As such, we do not include them in this work, but they can be incorporated into our computations once they are known.
However, we do have a code that computes first-order GSF in Kerr spacetime for both eccentric equatorial orbits~\cite{vandeMeent:2016pee} and generic orbits~\cite{vandeMeent:2017bcc} in the outgoing radiation gauge.
We have adapted our code to compute the first ever GSF results for spherical orbits.

In order to drive inspirals, we require a model for the GSF can be rapidly evaluated for any point in the parameter space. 
This is typically done by calculating the GSF at many points and then fitting or interpolating the resulting data~\cite{Warburton2012,Osburn2016}. 
We opt to interpolate our self-force data using Chebyshev polynomials, a technique which has proven to be very effective for EMRI calculations \cite{Lynch:2021ogr, Skoupy:2022adh}.
We grid the parameter space on a $18 \times 9$  Chebyshev nodes for prograde orbits where the primary spin is $0.9M$ and for inclination angles up to $45^\circ$ with respect to the equatorial plane, and interpolate the data using Chebyshev polynomials to sub-percent accuracy.

With an interpolated model in hand, we can then compute the inspiral trajectory using the method of osculating geodesics (OG)~\cite{Pound2008,Gair2011}.
We note that the OG equations for generic Kerr inspirals derived in either Ref.~\cite{Gair2011} or Ref.~\cite{Lynch:2021ogr} become singular in the spherical case.
This is not a physical singularity and in this work we derive OG equations in this limit that are finite (see Appendix~\ref{section:OG_Spherical}).  In this paper we explicitly assume the eccentricity is always zero so that the resulting inspiral model depends only on spherical-geodesic first-order GSF data. Neglecting terms proportional to eccentricity in our model incurs an error at 1PA order in the waveform phase but this is acceptable as we exclude other terms that contribute at this order, such as the currently unknown second-order GSF. The main goal for this work is not to produce a complete 1PA inspiral model, but to demonstrate how such a model can made computationally efficient without significant loss of accuracy.

Using OGs equations can take minutes to hours just to compute a single inspiral which is prohibitively slow for LISA data analysis.
This is due to the OG equations' dependence on the rapidly oscillating orbital phases, which forces a numerical solver to take small time steps to resolve all $\sim 10^4 - 10^7$  orbital cycles of a typical EMRI. 
To overcome this, we employ the technique of near identity (averaging) transformations (NITs) to find equations of motion that average out the dependence on the orbital phases while faithfully capturing the long term behaviour of the binary.
This framework was first sketched out for generic EMRI systems and applied to eccentric Schwarzschild inspirals in Ref.~\cite{vandeMeent:2018cgn}, hereafter paper I, and then extended to equatorial, eccentric Kerr inspirals in Ref.~\cite{Lynch:2021ogr}, hereafter paper II.
In both cases, the resulting inspirals retained sub-radian accuracy when compared to the OG equations and could be computed in less than a second.
However, in both of these works, the resulting inspirals were not parametrized in terms of Boyer-Lindquist coordinate time $t$ (the time of an asymptotic observer), which is much more convenient for data analysis as this is directly related to the time at the detector.
Overcoming this required an additional interpolation step which increases the overall computation time.
Following a procedure presented in Ref.~\cite{Pound2021}, one can parametrize the averaged equations of motion in terms of $t$ by performing an additional averaging transformation.
We apply this improved NIT to the case of quasi-spherical Kerr inspirals and find that the inspiral calculation retains sub-radian accuracy and sub-second computation time, whilst being parametrized in a form that is more convenient for practical waveform production.

Much of the work done to model EMRIs to post-adiabatic order makes use of the two-timescale framework~\cite{Hinderer2008,Miller2021,Pound2021} which has been long known to operate in an almost equivalent way to NITs~\cite{Kevorkian1987} when applied to the equations of motion.
Inspired by Ref.~\cite{Pound2021}, we perform a two-timescale expansion (TTE) on the NIT equations of motion and show that the TTE equations of motion also produces fast inspirals that are accurate to post-adiabatic order but with the drawback of doubling the number of equations to numerically solve.
On the other hand these equations are independent of the mass ratio and so, after being solved once (in under a second), inspirals for any mass ratio can be generated in a fraction of a second.
This could provide extra efficiency for an appropriately designed search or parameter estimation algorithm.~The TTE equations of motion can also be used in conjunction with the two timescale expansion of the field equations to produce complete post-adiabatic waveforms~\cite{Wardell:2021fyy}.

Unlike the OG equations, both the NIT and TTE equations of motion cleanly separate the adiabatic and post-adiabatic contributions. 
As such, it is straightforward to increase the overall accuracy of the inspiral by calculating the adiabatic contributions from computationally cheaper, and generally more accurate, gravitational wave flux calculations and interpolating them to higher precision. 
Since this is the leading order contribution to the inspiral evolution, it will need to be known with a fractional accuracy of $\sim \mr$~\cite{Osburn2016,Burke2021}.
We achieve this using a $18 \times 19$ Chebyshev grid which covers all inclinations and reaches a fractional accuracy in the adiabatic pieces of $\lesssim 10^{-5}$.
Using these higher precision interpolants for the adiabatic pieces along with the GSF model for the post-adiabatic pieces results in an improvement to the phase ranging from $10 - 10^4$ radians, highlighting just how important high precision flux calculations are to both adiabatic and post-adiabatic inspirals~\cite{Isoyama:2021jjd,Hughes2021}.

With our more accurate quasi-spherical inspiral model, we examine the effects of the self-force on quasi-spherical Kerr inspirals. 
We find that the evolution is consistent with what was known at adiabatic order and that the post-adiabatic contributions of the first-order GSF become more important as the inclination of the orbit increases.

This paper is organized as follows. 
In Section \ref{section:Spherical_Inspirals}, we introduce spherical geodesic orbits before giving an overview of the method of osculating geodesics.
In Section \ref{section:GSF}, we briefly review the gravitational self-force approach, discuss the modifications made to the Kerr GSF code of Ref.~\cite{vandeMeent:2017bcc} for spherical orbits and outline our interpolation scheme for both the gravitational wave fluxes and the GSF.
In Section \ref{section:Averaging_Techniques}, we review the technique of NITs for generic EMRI systems before describing an additional averaging transformation used to obtain averaged equations of motion parametrized by $t$. 
We then perform a two timescale expansion on these averaged equations of motion before applying these frameworks to the case of quasi-spherical inspirals.
In Section \ref{section:Implimentation}, we outline our numerical implementation for calculating and using these averaged equations of motion.
We demonstrate the results of this implementation in Section \ref{section:Implimentation} where we start by comparing inspirals and waveforms calculated using either OG, NIT or TTE equations of motion. 
We then show the effect of using higher precision adiabatic pieces from flux calculations before finally exploring the effects of the self-force across a subsection of the spherical Kerr parameter space.
We conclude with some discussion in Section \ref{section:Discussion}.
We detail how one can derive OG equations in the quasi-spherical limit in Appendix \ref{section:OG_Spherical} and how the phases used in the $t$ parametrized equations of motion relate to the waveform phases in Appendix \ref{section:tNIT_and_Wavefrom_Phases}.
 
 Throughout this paper we work in geometrical units where $G = c = 1$.

\section{Quasi-Spherical Inspirals around a Rotating Black Hole} \label{section:Spherical_Inspirals}
In this section we describe the motion of a non-spinning compact object of mass $\mu$ on a spherical orbit of Boyer-Lindquist radius  $\r$, with arbitrary inclination $\z = \cos \th$ in the Kerr spacetime under the influence of some arbitrary force.
We model the secondary as a point particle with a position given in modified Boyer-Lindquist coordinates by $\pos{x}^\alpha = \{\t,\r,\z,\ph \}$. Note that hereafter, we use a subscript `$p$' to denote a quantity evaluated at the particle's location.
Later in this work, we will take this perturbing force to be the self-force experienced by the secondary due to its interaction with its own metric perturbation. 
We denote the mass of the primary by $M$ and parametrize its spin by $a = |J|/M$ where $J$ is the angular momentum of the black hole.
The Kerr metric can then be written as
\begin{equation}\label{eq:metric}
	\begin{split}
		ds^2 = & - \left(1 - \frac{2 M r}{\Sigma} \right) dt^2 + \frac{\Sigma}{\Delta} dr^2 
		\\& + \frac{\Sigma}{1-z^2} dz^2 - \frac{4 M a r (1-z^2)}{\Sigma}dt d\phi
		\\& + \frac{1-z^2}{\Sigma} (2a^2r (1 - z^2) + \Sigma \varpi^2) d\phi^2 
	\end{split}
\end{equation}
where	$\Delta := r^2 + a^2 - 2 M r$, $\Sigma:= r^2 + a^2 z^2$, and $\varpi := \sqrt{r^2 + a^2}$.
If a force acts upon the secondary, its motion can be described by the forced geodesic equation
\begin{equation} \label{eq:forced_geodesic_eq}
	u^\beta\nabla_\beta u^\alpha = a^{\alpha}
\end{equation}
where $u^\alpha = dx^\alpha/d\tau$ is the secondary's four-velocity, $\nabla_\beta$ is the covariant derivative with respect to the Kerr metric, and $a^{\alpha}$ is the secondary's four-acceleration. 
We seek to recast Eq.~\eqref{eq:forced_geodesic_eq} into a form that is useful for applying the near-identity transformations.
Before considering the forced equation it is useful to first examine the geodesic limit.

\subsection{Geodesic motion and orbital parametrization}
In the absence of any perturbing force, the secondary will follow a geodesic, i.e.,
\begin{equation} \label{eq:geodesic_eq}
	u^\beta\nabla_\beta u^\alpha = 0.
\end{equation} 
The symmetries of the Kerr spacetime allow for the identification of integrals of motion given by
\begin{subequations}\label{eq:ELK}
	\begin{gather}
		\En = - u_t,   \quad   \Lz = u_\phi,  \tag{\theequation a-b}
		\\   \quad  \Q = u_z^2  +  \z^2 \left(\mathcal{A}+ \frac{\Lz^2}{1-\z^2} - u_z^2\right),  \tag{\theequation c}  
	\end{gather}
\end{subequations}
where $\En$ is the orbital energy per unit rest mass $\mu$, $\Lz$ is the z-component of the angular momentum divided by $\mu$, $\Q$ is the Carter constant divided by $\mu^2$ ~\cite{Carter1968}, and $\mathcal{A} = a^2 (1-\En^2)$.
This definition of the Carter constant is related to another common definition of the Carter constant, $\K$, by 
\begin{equation}\label{eq:Carter_Constant}
	\K = \Q + (\Lz - a \En)^2.
\end{equation}
The geodesic equation can be written explicitly in terms of these integrals of motion~\cite{Carter1968,Drasco2004}:
\begin{subequations}\label{eq:Geodesic_eqs}
	\begin{align}
		\begin{split} \label{eq:Vr}
			\left( \frac{d \r}{d\lambda} \right)^2  \coloneqq V_r  &= \pos{\mathcal{B}}^2 - \pos{\Delta} \left(\r^2+ \K \right) 
			   = 0  
		\end{split}\\
		\begin{split}\label{eq:Vz}
			\left( \frac{d \z}{d\lambda} \right)^2  \coloneqq V_z &= \Q - \z^2 \left(\mathcal{A} (1-\z^2) + \Lz^2 + \Q \right) \\
			&= (\z^2-z_-^2)\left(\mathcal{A} \z^2 - z_+^2 \right)   
		\end{split}\\
		\begin{split}\label{eq:Geodesic_t}
			\frac{d \t}{d\lambda}  \coloneqq f_t^{(0)} &= \frac{\pos{\varpi}^2}{\pos{\Delta}} \pos{\mathcal{B}} - a^2 \En(1-\z^2)+ a \Lz 
		\end{split}\\
		\begin{split} \label{eq:Geodesic_phi}
			\frac{d \ph}{d\lambda}  \coloneqq f_\phi^{(0)} &= \frac{a}{\pos{\Delta}} \pos{\mathcal{B}} + \frac{\Lz}{1-\z^2} - a \En.   
		\end{split}
	\end{align}
\end{subequations}
 where $\mathcal{B} = \En \varpi^2 - a \Lz $, the radial potential $V_r = 0$ for spherical and circular orbits, and $z_+ > z_-$ are the roots of the polar potential $V_z$.
We make use of (Carter-)Mino time $\lambda$ which is related to the secondary's proper time $\tau$, by $d\tau / d \lambda = \pos{\Sigma}$, as this decouples the radial and polar equations ~\cite{Mino:2003yg}.
While this is not strictly necessary in the spherical case as there is no radial motion, using $\lambda$ as our time parameter allows us to take advantage of the analytic solutions for generic Kerr geodesics derived in Ref.~\cite{Fujita2009a}, and allows for a more uniform treatment with paper II and the generic case in the future. 

Rather than parametrize an orbit using the integrals of motion $\vec{\mathcal{J}} = \{ \En, \Lz, \K \}$, we find it more convenient to work with the constants $\vec{P} = \{\r, e = 0, x\}$.
Here $x$  is a measure of the orbital inclination given by $x = \cos \theta_{\text{inc}}$. 
The inclination angle $\theta_{\text{inc}}$ is related to $\theta_{\text{min}}$ (the minimum value of $\th$ measured with respect to the primary's spin axis) by $\theta_{\text{inc}} = \pi / 2 - \text{sgn}(\Lz) \theta_{\text{min}}$.
Explicit relationships between the integrals of motion $\vec{\mathcal{J}}$ in terms of  $\vec{P}$ for spherical orbits are too lengthy to be expressed here, but they were first derived in Ref.~\cite{Hughes2000} and can be found in the \texttt{KerrGeodesics} package as part of the Black Hole Perturbation Toolkit~\cite{BHPToolkit}. 

There are other common choices for inclination in the literature such as the inclination angle  $\iota$~\cite{Ryan:1995zm,Ryan:1995xi,Hughes2000,Hughes2001} given by $\cos \iota = \Lz/ \sqrt{\Lz^2 + \Q} $.
Both inclination angles smoothly parametrize all inclinations between prograde equatorial motion where $x = 1 =\cos \iota$ to retrograde equatorial motion where $x= -1 = \cos \iota$.
However, we opt to use $x$ as it has a simple relation to the roots of the polar potential $V_z$ via
\begin{align} \label{eq:polar_roots}
	\begin{split}
		z_- &= \sqrt{1-x^2}
	\end{split}\\
	\begin{split}
		z_+ &= \sqrt{\mathcal{A} + \Lz^2/x^2}.
	\end{split}
	\end{align}
It is worth noting that not all values of $\{\r,x\}$ correspond to bound geodesics and we denote the position of the innermost stable spherical orbit (ISSO) as $r_{\text{ISSO}}(a,x)$~\cite{Glampedakis2002,Stein2020}.

In order to later apply the near-identity (averaging) transformations, it will be useful to employ action-angle formulation to parametrize the geodesic motion~\cite{Schmidt2002,Fujita2009a,vandeMeent:2019cam}.
In this description the orbital phase $\vec{q} = \{q_z\}$ is such that the geodesic equations can written in the form
\begin{subequations}\label{eq:GeodesicEqsPandq}
	\begin{gather}
		\frac{d P_j}{d \lambda} = 0\quad \text{and}  \quad \frac{d q_z}{d \lambda} =  \Upsilon_z^{(0)} (\vec{P}),   \tag{\theequation a-b}
	\end{gather}
\end{subequations}
where $\Upsilon_z$ is the Mino time fundamental polar frequency. 
This is known analytically~\cite{Fujita2009a} and is given by
\begin{equation}
	\Upsilon_z^{(0)} = \frac{\pi z_+}{2 K(k_z) }
\end{equation}
where $k_z= \mathcal{A} z_-^2/z_+^2$, $K$ is the complete elliptic integral of the first kind with $K(m) = F(\pi/2 |m)$ and $F$ is the incomplete elliptic integral of the first kind given by
\begin{equation}
	F(\phi | m) = \int^\phi_0 \frac{d\theta}{\sqrt{1 - m \sin^2 \theta}}.
\end{equation}
One can also analytically express the polar coordinate $z$ in terms of $r,x$ and $q_z$ via 
\begin{equation} \label{eq:z_analytic}
	\z(q_z) = z_-\sn\left(K(k_z)\frac{2(q_z+\frac{\pi}{2})}{\pi} \bigg| k_z\right),
\end{equation}
where $\sn$ is the Jacobi elliptic sine function given by $\sn(u|m) = \sin(\am(u|m)) $ and the Jacobi amplitude $\am$ is the inverse function of $F$. 

At this point we note that it is also common in the literature~\cite{Drasco:2003ky, Nasipak:2019hxh} to express this polar coordinate in terms of a quasi-Keplerian angle $\chi_z$, via
\begin{equation}\label{eq:Keplerian_Angles}
		  \quad \z(\chi_z) = z_- \cos (\chi_z).
\end{equation}
However, the rate of change of $\chi_z$ depends on $\chi_z$ itself, which will be inconvenient in later sections as we look to derive averaged equations of motion that have no dependence on the orbital phases.

\subsection{Osculating Geodesics for quasi-spherical inspirals}
To go beyond the geodesic orbits and describe quasi-spherical inspirals, we make use of the method of osculating orbital elements (or osculating geodesics)~\cite{Pound2008,Gair2011} to recast the equations of motion of a body under the influence of a force obeying Eq.~\eqref{eq:forced_geodesic_eq}. 

We chose a set of parameters that uniquely specify a  geodesic orbit, such as the integrals of motion $\vec{P}$ along with the initial values of the orbital phases of the geodesic orbit $\vec{q}_0$, and designate them as a set of ``orbital elements" $\vec{I} = \{\vec{P},\vec{q}_0\}$ \ \footnote{Note that for inspiral trajectories, the orbital elements $\vec{q}_0(\lambda)$ are different quantities from the values of $\vec{q}$ evaluated at $\lambda = 0$, i.e., $\vec{q}(0)$.}. For spherical geodesics, we chose $\vec{I} = (r_p,x,q_{z,0})$. For accelerated orbits, these orbital elements are promoted from constants to functions of time.

 Here we  are implicitly assuming that the eccentricity $e$ stays zero throughout the inspiral of an initially spherical system. While this has been shown to be valid at 0PA order~\cite{Kennefick:1995za,Ryan:1995xi}, at 1PA this becomes a subtle issue~\cite{Loutrel:2018ssg,Will:2019lfe}. Whether the eccentricity remains zero throughout the post-adiabatic evolution is intimately related to the exact setup for the calculation of other post-adiabatic corrections such as the second-order fluxes~\cite{Warburton2021}. For the sake of simplicity, and to have an inspiral model that depends only on spherical-geodesic first-order GSF data, in this paper we explicitly assume the eccentricity is always zero. In doing so, we are neglecting terms that contribute at the same order as the currently unknown second-order GSF.

To produce waveforms, we will also require the evolution of certain ``extrinsic quantities" $\vec{S} = \{\pos{t},\pos{\phi}\}$ which are the secondary's Boyer-Lindquist time and azimuthal coordinates respectively.
The combined osculating geodesic evolution equations take the following form:
\begin{subequations}\label{eq:Generic_EMRI_EoM}
	\begin{align}
		\begin{split}
			\frac{d P_j}{d \lambda} &= F_j (\vec{P}, \vec{q}) ,
		\end{split}\\
		\begin{split}
			\frac{d q_i}{d \lambda} &=  \Upsilon^{(0)}_i (\vec{P}) +  f_i (\vec{P}, \vec{q}),
		\end{split}\\
		\begin{split}
			\frac{d S_k}{d \lambda} &= f_k^{(0)}(\vec{P}, \vec{q})
		\end{split}
	\end{align}
\end{subequations}
 where $f_i = d q_{0,i}/ d \lambda$.

For generic Kerr inspirals, this was implemented in Ref.~\cite{Gair2011} using quasi-Keplerian angles (e.g., $\chi$) as the orbital phases.
In order to derive averaged equations of motion, we implemented an alternative version in Paper~II~\cite{Lynch:2021ogr} which makes use of geodesic action angles $\vec{q}$ for the orbital phases and the integrals of motion.
In Appendix~\ref{section:OG_Spherical}, we derive the explicit expressions for spherical inspirals.
Combining these modified evolution equations with the rest of the evolution equations derived in Paper~II~\cite{Lynch:2021ogr} gives us equations of motion for quasi-spherical inspirals under the influence of the gravitational self-force with the form
\begin{subequations}\label{eq:Spherical_EMRI_EoM}
	\begin{align}
			\frac{d \r}{d \lambda} &= \mr F_{r}^{(1)}(\r,x,q_z) + \mr ^2  F_{r}^{(2)}(\r,x,q_z) + \HOT{3},\\
			\frac{d x}{d \lambda} &= \mr F_x^{(1)}(\r,x,q_z) + \mr^2 F_x^{(2)}(\r,x,q_z) + \HOT{3},\!\\
			\frac{d q_z}{d \lambda} &=  \Upsilon^{(0)}_z (\r,x) + \mr f^{(1)}_z (\r,x,q_z) + \HOT{2},\\
			\frac{d t}{d \lambda} &= f_t^{(0)}(\r,x,q_z),\\
			\frac{d \phi}{d \lambda} &= f_\phi^{(0)}(\r,x,q_z).
	\end{align}
\end{subequations}

\section{Gravitational self-force for quasi-spherical Kerr inspirals} \label{section:GSF}

The gravitational self-force approach consists of expanding the metric of the binary around the metric of the primary, i.e., $g_{\mu\nu} = \bar{g}_{\mu\nu} + \epsilon h^{(1)}_{\mu\nu} +\epsilon^2 h^{(2)}_{\mu\nu} + \dots$ where $\bar{g}_{\mu\nu}$ is the Kerr metric and the $h^{(n)}_{\mu\nu}$ are the \mbox{$n$-th} order perturbations to the spacetime due to the presence of the secondary. 
Using matched asymptotic expansions, one can then derive how the interaction between these metric perturbations and their source effects the motion of the secondary~\cite{Poisson:2011nh}.

\subsection{Gravitational self-force for spherical orbits}

This work represents the first calculation of the gravitational self-force for spherical orbits in Kerr spacetime. While the gravitational self-force had been previously calculated for generic (eccentric and inclined) orbits in Kerr in~\cite{vandeMeent:2017bcc}, building on previous developments in~\cite{Shah:2012gu,Pound:2013faa,vandeMeent:2015lxa,vandeMeent:2016pee}, and the scalar self-force for spherical orbits had been calculated in~\cite{Warburton:2014bya}, the case of the gravitational self-force for spherical orbits has not been explored before now.

To calculate the gravitational self-force for spherical orbits, we follow the same method as was used in the generic orbit case in~\cite{vandeMeent:2017bcc}, which can be adapted without major adjustments.
We summarize our method here. We start by solving the Teukolsky equation for the Weyl scalar $\psi_4$ in the frequency domain using a numerical implementation~\cite{Fujita:2004rb,Fujita:2009us} of the semi-analytical Mano-Suzuki-Takasugi method~\cite{Mano:1996vt,Mano:1996gn}. 
From $\psi_4$, we obtain the Hertz potential by algebraically inverting the fourth order differential equation relating it to $\psi_4$~\cite{Ori:2002uv,vandeMeent:2015lxa} mode-by-mode. 
From the modes of the Hertz potential we can reconstruct the modes of the local metric perturbation in the outgoing radiation gauge~\cite{Chrzanowski:1975wv,Kegeles:1979an, Wald:1978vm}, from which, in turn, we obtain the modes of the gravitational self-force.
Initially these modes are expressed in spin-weighted spheroidal harmonics and their derivatives. 
To facilitate calculating the regular part of the GSF modes using mode-sum regularization~\cite{Barack:1999wf,Barack:2001gx,Barack:2002mh,Barack:2009ux}, the spheroidal harmonics are first projected to spin-weighted spherical harmonics using the method introduced in~\cite{Hughes2000}, which in turn are projected onto scalar spherical harmonics~\cite{vandeMeent:2016pee,vandeMeent:2017bcc,vandeMeent:2015lxa}. 
The mode-sum is accelerated by a tail fitting procedure leading to an exponential convergence of the mode-sum~\cite{vandeMeent:2015lxa}.

The above metric reconstruction procedure captures most of the metric perturbation, except for a perturbation of the background Kerr spacetime's mass and spin, and a purely gauge contribution. 
The mass and spin completions are directly related to the energy and angular momentum of the orbit~\cite{Merlin:2016boc,vanDeMeent:2017oet}.  
The gauge completion is more involved. 
The version of the outgoing radiation gauge used for the GSF calculation, the no-string gauge~\cite{Pound:2013faa}, has a distributional singularity on sphere of the orbit. 
Recently,~\cite{Toomani:2021jlo}~has proposed a generalization of our method that produces the metric perturbation in a gauge that is sufficiently smooth to allow it to be used as a source for a second order calculation. 
We here strive for a less lofty goal and will add the minimal gauge completion needed to ensure orbit averaged observables like the frequency-shifts are invariant. 
This involves fixing a 2-parameter gauge freedom representing a rescaling of time and time dependent coordinate rotations inside the orbit. 
In~\cite{Toomani:2021jlo} a nice derivation of the needed coefficients was given. However, for this work we used an implementation~\cite{gaugecompletion} of a generalization of the method used for the mass and spin completion in Ref.~\cite{Merlin:2016boc}. 
The gauge vector representing this freedom is given by,
\begin{equation}
\xi^\mu = \{\alpha t, 0, 0, \beta t\},
\end{equation}
with the gauge fixing parameters $\alpha$ and $\beta$ given by
\begin{align}
\label{eq:gaugecompinta}
\alpha &= \frac{1}{2\pi \Upsilon_t}\int_{0}^{2\pi}  \frac{\mathcal{I}_\alpha(\z)}{3 \Delta^2_p \Sigma^2_p(\r^2+a^2)(1-\z^2)^2}  dq_z,\text{ and}
\\
\label{eq:gaugecompintb}
\beta &= \frac{1}{2\pi \Upsilon_t}\int_{0}^{2\pi}  \frac{\mathcal{I}_\beta(\z)}{3 a \r \Delta^2_p \Sigma^2_p(\r^2+a^2)(1-\z^2)^2}  dq_z,
\end{align}
where the numerators of the integrands, $\mathcal{I}_\alpha$ and $\mathcal{I}_\beta$ are given in Appendix~\ref{app:gaugecomp}, and  $\Delta_p$ and  $\Sigma_p$ denote $\Delta$ and $\Sigma$ evaluated at the particle location.

It is useful to split the self-force into dissipative (time anti-symmetric) and conservative (time-symmetric) contributions~\cite{Barack:2009ux}.
The dissipative pieces cause the orbit to shrink until the secondary plunges into the primary.
They also have a small effect on the orbital inclination, causing the orbit to become more inclined over time~\cite{Ryan:1995xi,Ryan:1995zm,Hughes2000,Hughes2001}.
To produce adiabatic waveforms, we only require knowledge of the orbit averaged dissipative pieces of the first-order self-force.
These can be related, via balance laws, to the fluxes of GWs to infinity and down the event horizon.
Since calculating fluxes avoids regularization of the metric perturbation, adiabatic inspirals are typically calculated via flux balance laws~\cite{Glampedakis2002,Hughes2005,Fujita:2020zxe,Isoyama:2021jjd, Hughes2021}.
The conservative pieces have more subtle effects on the inspiral, such as altering the rate of periapsis advance, the rate of nodal precession, and the location of the innermost stable circular orbit~\cite{Barack2009,Barack2010a,Barack2011,Warburton2011,vandeMeent:2016hel, Vines:2015efa, Fujita:2016igj}.

Computing post-adiabatic inspirals requires knowledge of both the dissipative and conservatives pieces of the first-order self-force and the orbit average piece of the second-order self-force~\cite{Hinderer2008}.
There are as yet no calculations of the latter in Kerr spacetime, so we will make do with only the first-order gravitational wave fluxes and the first-order self-force in this work, though this means that the resulting ``post-adiabatic" inspirals and waveforms will be gauge dependent until the missing second-order contributions are added~\cite{Lynch:2021ogr}.

In order to drive the  OG equations of motion to calculate the inspiral, we require a model for the GSF that can be rapidly evaluated at each time step during the inspiral.
This is typically done by tiling the parameter space with with GSF data which is then either fitted to a model or interpolated.
While this has been done in a variety of ways for eccentric orbits in Schwarzschild~\cite{Warburton2012, Osburn2016} and Kerr~\cite{Lynch:2021ogr}, we will now describe the first interpolated GSF model for quasi-spherical Kerr inspirals.

\subsection{Interpolated gravitational wave fluxes for quasi-spherical Kerr inspirals}
To introduce our interpolation procedure, we will first interpolate the energy and angular momentum fluxes for spherical orbits.
Since flux calculations, which can be obtained directly from $\psi_4$, are significantly cheaper than calculating the GSF, it is much more feasible to densely tile a large section of the parameter space with flux data.
This in turn will result in more accurate interpolation of the leading-order, adiabatic effects.
It will also allow us to carry out consistency checks on our GSF model by comparing the fluxes with the orbit averaged GSF.

We start by fixing the value of the spin parameter of the primary, which we choose to be $a = 0.9M$.
This reduces our parameter space to two parameters; the orbital radius $\r$ and the inclination $x$. 
We then define a parameter $y$ using $\r$ and the position of the innermost stable spherical orbit $r_{\text{ISSO}}$. 
We choose $y$ to be 
\begin{equation}\label{eq:y_Kerr}
	y = \sqrt{\frac{r_{\text{ISSO}}(a,x)}{\r{}}}.
\end{equation}
Tiling the parameter space with $y$ instead of $\r$ will concentrate more points near the ISSO where the fluxes and the GSF experience the most variation.
We let $y$ range from $0$ to $1$ and $x$ range from $-1$ to $1$ thus covering all inclinations for both prograde and retrograde orbits.

This parametrization is convenient when using Chebyshev polynomials of the first kind, where the order $n$ polynomial is defined by $T_n (\cos \vartheta) \coloneqq \cos (n \vartheta)$.
The Chebyshev nodes are the roots these polynomials, and the location of the $k$th root of $n$th polynomial is given by
\begin{equation}
	N(n,k)= \cos{\left( \frac{2 k -1}{2n} \pi \right)}
\end{equation}
We then calculate the  fluxes  on a $18 \times 19$ grid of Chebyshev nodes, with the $y$ values given by the roots of the 18th order Chebyshev polynomial and the $x$ values given by the roots of the 19th order Chebyshev polynomial. The size of this grid was chosen empirically until the interpolant reached the desired accuracy across the parameter space.

At each point on this grid, we calculate the energy and angular momentum flux both at infinity and at the event horizon. 
From these, one can calculate the leading order orbit averaged rate of change of the energy and angular momentum of the secondary via the following balance laws \cite{Galtsov:1982hwm,Quinn:1999kj,Mino:2003yg, Mino:2005an, Mino:2005yw}:
\begin{subequations} \label{eq:flux_balance}
	\begin{align}
		\avg{\frac{d \En}{d t}} &= - \mr (\mathcal{F}^\infty_\En + \mathcal{F}_\En^\mathcal{H}) + \HOT{2}, \\
		 \avg{\frac{d \Lz }{d t}}& = - \mr (\mathcal{F}_\Lz^\infty + \mathcal{F}_\Lz^\mathcal{H}) + \HOT{2},
	\end{align}
\end{subequations}
where the orbit average of a quantity $A$ is given by
\begin{equation}\label{eq:average}
	\avg{A} (\vec{P}) =  \frac{1}{2\pi} \int^{2 \pi}_{0} A(\vec{P},q_z) d q_z.
\end{equation}

We could interpolate the rates of change of energy and angular momentum, but we find it more convenient to work with the orbital elements $\r$ and $x$.
As such, we find their rates of change via the chain rule:
\begin{subequations} \label{eq:flux_chain_rule}
	\begin{align}
		\avg{\frac{d \r}{d t}} &= \frac{\partial \r}{\partial \En} \avg{ \frac{d\En}{d t}} + \frac{\partial \r}{\partial \Lz} \avg{ \frac{d\Lz}{d t}} + \HOT{2} \\
		 &= \mr \Gamma_r^{(1)} + \HOT{2} \nonumber
\\
		\avg{\frac{d x}{d t}} &= \frac{\partial x}{\partial \En} \avg{ \frac{d\En}{d t}} + \frac{\partial x}{\partial \Lz} \avg{ \frac{d\Lz}{d t}} + \HOT{2}\\\
		&= \mr \Gamma_x^{(1)} + \HOT{2}.\nonumber
	\end{align}
\end{subequations}
The partial derivatives can be found using the analytic expressions for $\En(\r,x)$ and $\Lz(\r,x)$ from Ref.~\cite{Hughes2000} to construct the Jacobian
\begin{equation}
	J = 
	\begin{bmatrix}
		\frac{d \En}{d \r} & \frac{d \En}{d x}\\
		\frac{d \Lz}{d \r} & \frac{d \Lz}{d x}
	\end{bmatrix}.
\end{equation}
This can then be inverted to give
\begin{equation}
	J^{-1} = 
	\begin{bmatrix}
		\frac{d \r}{d \En} & \frac{d \r}{d \Lz}\\
		\frac{d x}{d \En} & \frac{d x}{d \Lz}
	\end{bmatrix}
	= \frac{1}{\det{J}}
	\begin{bmatrix}
		\frac{d \Lz}{d x} & -\frac{d \En}{d x}\\
		-\frac{d \Lz}{d \r} & \frac{d \En}{d \r}
	\end{bmatrix}.
\end{equation}

To improve the accuracy of our interpolation, we rescale the data for $\Gamma_r^{(1)}$ and $\Gamma_x^{(1)}$ by a factor of $\r^3(1-y)$ and $\r^{11/2}(1-x^2)$ respectively.
This scaling comes from the leading order PN terms for $\Gamma_r^{(1)}$ and $\Gamma_x^{(1)}$, times a term that is zero for the limiting cases of either the separatrix or the equatorial plane respectively.
Finally, we take a discrete cosine transformation of the data on our Chebyshev grid to obtain the Chebyshev polynomial coefficients $C_{\r/x}^{ij}$.
Summing these coefficients together with Chebyshev polynomials gives us the following interpolants for $\Gamma_r^{(1)}$ and $\Gamma_x^{(1)}$:

\begin{subequations}
	\begin{equation}
		\Gamma_{r}^{(1)} =\frac{1}{\r^3(1-y)} \sum_{i = 0}^{17}  \sum_{j = 0}^{18} C^{i j}_{p} T_i \left(2y-1\right) T_j\left(x\right),
	\end{equation}
	\begin{equation}
		\Gamma_x^{(1)} =\frac{1}{\r^{11/2}(1-x^2)} \sum_{i = 0}^{17}  \sum_{j = 0}^{18} C^{i j}_{x} T_i \left(2y-1\right) T_j\left(x\right).\!\!
	\end{equation}
\end{subequations}
Using the largest coefficient for $i = 17$ and $j =18$ to estimate the relative error, we infer that these interpolants should have a relative error of $\sim 10^{-6}$.

 \begin{figure}
	%\centering
	\subfloat[Relative error of the energy flux. \label{fig:FluxEInterpolation} ]{
		\includegraphics[width=0.49\textwidth]{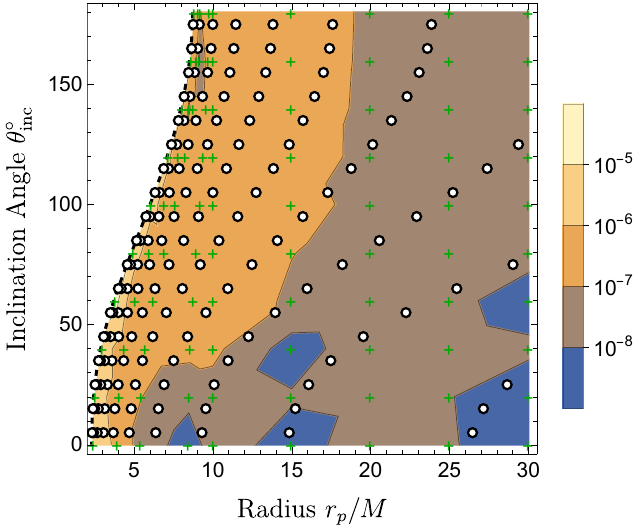}
	}
	\hfill
	\subfloat[Relative error of the angular momentum flux. \label{fig:FluxLInterpolation} ]{
		\includegraphics[width=0.49\textwidth]{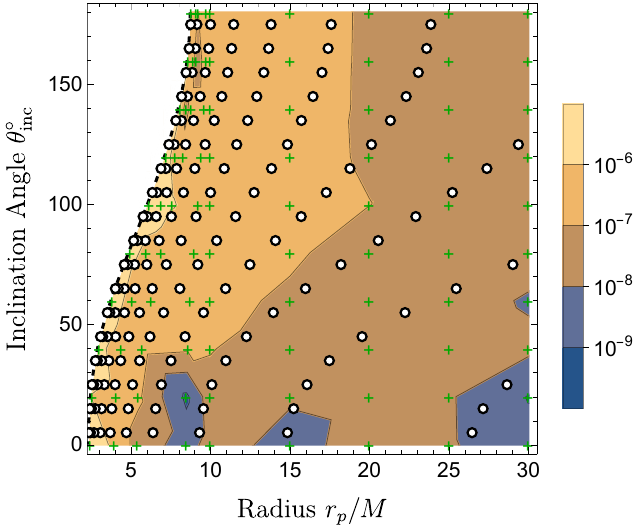}
	}
	\hfill
	\caption{Relative error in the rate of energy (a) and angular momentum (b) loss between the interpolated fluxes and the fluxes calculated with a grid of verification points that were not used for the interpolation. The white dots indicate the locations of the data points that were interpolated to produce the model and the green crosses indicate the positions of verification data. The relative error is always $<10^{-5}$ with the exception of orbits that are very close to the ISSO.} \label{fig:FluxInterpolationError}
\end{figure}
To test the accuracy of this interpolation, we compare the interpolants to data on a grid that were not used for the interpolation while using the relations $\mathcal{F}_\En ^ \infty + \mathcal{F}_\En^\mathcal{H} = - \left( \partial \En / \partial p \right) \Gamma^{(1)}_p - \left( \partial \En / \partial x \right) \Gamma^{(1)}_x$ and $\mathcal{F}_\Lz^ \infty + \mathcal{F}_\Lz^\mathcal{H}  = - \left( \partial \Lz / \partial p \right) \Gamma^{(1)}_p - \left( \partial \Lz / \partial  x \right)  \Gamma^{(1)}_x$.
In Fig.\ref{fig:FluxInterpolationError}, we see that the interpolants match the energy and angular momentum fluxes to a relative error of $ \lesssim 10^{-5}$ with the exception of points very close to the ISSO.
In principle this model interpolates out to $\r=\infty$ and the leading PN scaling should give it the right behaviour in the weak field. We would expect the model to retain a comparable level of accuracy at large $\r$, but it has not been tested beyond $\r \sim 30$ as, in this work, we are more interested in the strong field dynamics.

\subsection{Interpolated gravitational self-force model for quasi-spherical Kerr inspirals}

We now use a similar interpolation scheme to create a model for the gravitational self-force that is continuous throughout the parameter space and fast to evaluate which can be used with the osculating geodesic equations to describe quasi-spherical Kerr inspirals.

However, given the cost of our GSF code ---the computation of the GSF on a single geodesic can take hundreds of CPU hours --- we restrict ourselves to a 2D slice of the EMRI parameter space using only a few hundred points.
Once again, we restrict $a = 0.9 M$ and let $y$ range from $y_{\text{min}} = 0.1$ to $y_{\text{max}} = 1$, but instead of $x$, we opt to tile in $z_-^2 = 1-x^2$ and let $z_{-,\text{min}}^2 = 0$ to $z_{-,\text{max}}^2 = 0.5$. This allows us to cover inclination angles ranging from $0^\circ \leq \theta_\text{inc} \leq 45^\circ $, in a way that is easy to rescale for Chebyshev interpolation.

We define parameters $u$ and $v$ which cover this parameter space as they range from $(-1,1)$
\begin{subequations}
	\begin{equation}
		u \coloneqq \frac{y-(y_{\text{min}}+y_{\text{max}})/2}{(y_{\text{min}}-y_{\text{max}})/2} \text{ and},
	\end{equation}  
	\begin{equation}
		v \coloneqq \frac{z_-^2-(z_{-,\text{min}}^2+z_{-,\text{max}}^2)/2}{(z_{-,\text{min}}^2-z_{-,\text{max}}^2)/2}.
	\end{equation}
\end{subequations}

We then calculate the GSF on a $18 \times 9$ grid of Chebyshev nodes, with the $u$ values given by the roots of the 18th order polynomial and the $v$ values given by the roots of the 9th order polynomial.
At each point on our grid, we Fourier decompose each component of the force with respect to the polar action angle $q_z$. As inclination increases more and more Fourier modes of the GSF become relevant. However, the number of relevant modes stays finite even for polar orbits. To build a Chebyshev interpolant we need to resolve all Fourier modes at all grid points. Resolving the higher order harmonics at low inclination grid points is numerically challenging and therefore computationally expensive. This is the main reason for limiting the range of inclinations to $0^\circ \leq \theta_\text{inc} \leq 45^\circ $ as this limits the number of Fourier modes that need to be resolved.

To smooth the behaviour of the force near the separatrix and improve the accuracy of our interpolation, we then multiply the data for each Fourier coefficient by a factor of $(1-y)^2$.
Next, we use Chebyshev polynomials to interpolate each Fourier coefficient across the $(u,v)$ grid. 
We then sum the modes to reconstruct our interpolated gravitational self-force model:
\begin{subequations}
	\begin{equation}
		a_\mu =  \sum_{\kappa = 0}^{24} \frac{ A^\kappa_\mu(y,z_-^2) \cos (\kappa q_z) + B^\kappa_\mu(y,z_-^2) \sin (\kappa q_z)}{(1-y)^2},
	\end{equation}
	\begin{equation}
		A^\kappa_\mu(y,e) = \sum_{i = 0}^{17}  \sum_{j = 0}^{8} A^{\kappa i j}_{\mu} T_i \left(u\right) T_j\left(v\right),
	\end{equation}
	\begin{equation}
		B^\kappa_\mu(y,e) = \sum_{i = 0}^{17}  \sum_{j = 0}^{8} B^{\kappa i j}_{\mu} T_i \left(u\right) T_j\left(v\right).
	\end{equation}
\end{subequations}

We note that this choice of rescaling forces each component to become singular at the ISSO, and while the components of the GSF change rapidly as one approaches the ISSO, we still expect them to be finite at the ISSO.
A greater understanding of the analytic structure of the GSF in this region would greatly improve this and any future interpolated GSF models.

We also note that the GSF should satisfy the orthogonality condition with the geodesic four-velocity, i.e., $a_{\mu} u^{\mu} = 0$.
Interpolation will bring with it a certain amount of error which can cause this condition to be violated. 
Since the OG equations are derived assuming this condition to be true~\cite{Gair2011,Lynch:2021ogr}, we project the force so that this condition is always satisfied, i.e.,
\begin{equation}
	a_\mu^\perp = a_\mu + a_\nu u^\nu u_{\mu}.
\end{equation}

To verify the accuracy of our interpolated model, we employ the flux balance laws to compare the local energy and angular momentum lost by the secondary against the energy and angular momentum fluxes radiated at infinity and down the horizon:
\begin{subequations}\label{eq:FluxBalance}
	\begin{align}
	\begin{split}
		\frac{\Upsilon_\tau^{(0)}}{\Upsilon_t^{(0)}} \avg{a_t^\perp} &= \left(\mathcal{F}_{\En,\infty} + \mathcal{F}_{\En,H}\right),
	\end{split}\\
	\begin{split}\\
		\frac{\Upsilon_\tau^{(0)}}{\Upsilon_t^{(0)}} \avg{a_\phi ^\perp} &= -\left(\mathcal{F}_{\Lz,\infty} + \mathcal{F}_{\Lz,H}\right),
	\end{split}
	\end{align}
\end{subequations}
where $\Upsilon_t^{(0)} = \avg{f_t^{(0)}}$ and $\Upsilon_\tau^{(0)} = \avg{\pos{\Sigma}}$  are the fundamental Mino time frequencies for the secondary's Boyer-Lindquist time coordinate and proper time with analytic expressions found in Ref.~\cite{Fujita2009a} and appendix C of Ref.~\cite{vandeMeent:2019cam} respectively.

 \begin{figure}
	%\centering
	\subfloat[Relative error for rate of change of energy. \label{fig:FtvsEFlux} ]{
		\includegraphics[width=0.49\textwidth]{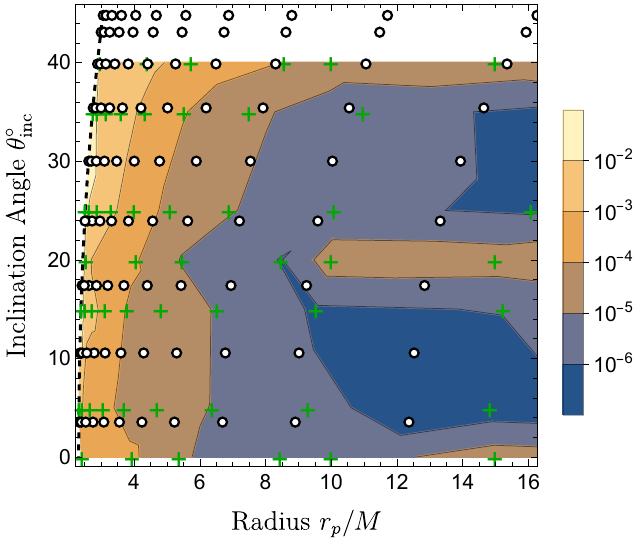}
	}
	\hfill
	\subfloat[Relative error for rate of change of angular momentum. \label{fig:FphivsLFlux} ]{
		\includegraphics[width=0.49\textwidth]{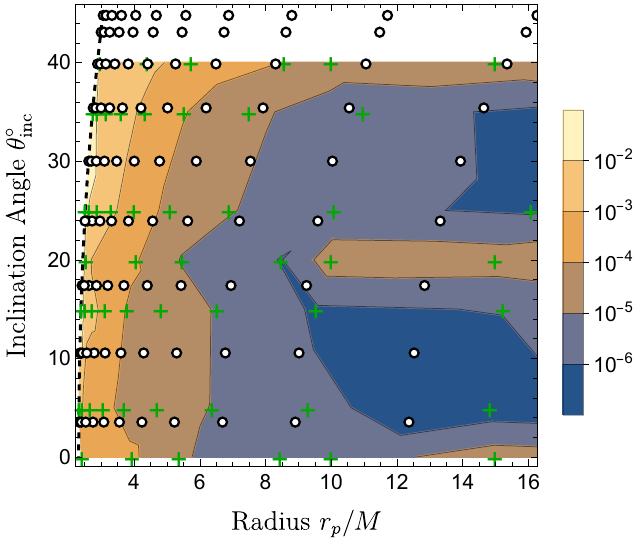}
	}
	\hfill
	\caption{
		Relative error in the rate of energy (a) and angular momentum (b) loss between the asymptotic fluxes and the interpolated GSF model. 
		The white dots indicate the locations of the data points the model is interpolating. 
		The green crosses indicate the locations of the flux calculations and thus where the comparisons are made. 
		The relative error only rises to $\sim10^{-2}$ when close to the ISSO, and is otherwise typically of the order $\sim 10^{-3} - 10^{-5}$. \label{fig:InterpolationError}}
\end{figure}

Fig.~\ref{fig:InterpolationError} shows the relative error in the fluxes across the parameter space.
The white circles indicate the points used to generate the interpolated GSF model, whereas the green crosses indicate the points in the parameter space where the fluxes were calculated and thus where the comparisons were made.
We find that the relative error in the fluxes only rises to $\sim 10^{-2}$ when close to the ISSO, and it is typically of the order $\sim 10^{-3} - 10^{-5}$. 
This is comparable to previous methods which required tens of thousands of points to achieve the same level of accuracy~\cite{Warburton2012,Osburn2016}.
While this would not be sufficiently accurate for LISA data analysis which would require fluxes accurate to $\sim \mr$, such an interpolating error may be permissible for the post-adiabatic contributions of the first order GSF~\cite{Osburn2016, Burke2021}.
We will go on to improve this model by incorporating the higher precision flux interpolants in Sec.~\ref{section:Flux_GSF_Hybrid}.

We can now use this model in conjunction with the OG equations of motion~\eqref{eq:Spherical_EMRI_EoM} to calculate inspiral trajectories.
However, we find these trajectories take minutes to hours to compute, due to the need to resolve hundreds of thousands of orbital cycles. 
We will now look to leverage averaging transformations which can remove the dependence of the orbital phases from the equations of motion while retaining the accuracy required to produce post-adiabatic EMRI waveforms.

\section{Averaging techniques for EMRI equations of motion} \label{section:Averaging_Techniques}

NITs are a well known technique in applied mathematics and celestial mechanics~\cite{Kevorkian1987}.
This approach involves making small transformations to the equations of motion, such that the short timescale physics is averaged over, while retaining information about the long-term evolution of a system.
In paper I~\cite{vandeMeent:2018cgn}, these transformations were derived for a generic EMRI system which we review in Section~\ref{section:NIT_Review} below. 

As we have seen above, for Kerr geodesics it is helpful to parametrize the motion by Mino time, $\lambda$.
However, for data analysis one requires the final waveforms to be parametrized by the time at the detector.
Naively, transforming back to Boyer-Lindquist time, $t$, requires either numerically interpolating $\lambda (t)$, or adding orbital timescale oscillations back into our NIT equations of motion. 
The former additional step is unwelcome in the pursuit of fast waveform generation, and the latter would remove all the benefit of the averaging transformation.
Fortunately, it was noted in Ref.~\cite{Pound2021} that these oscillations can be averaged out if one performs an additional averaging transformation. 
We outline the details of this procedure in Section~\ref{section:t_NIT}.

NITs are not the only averaging procedure that can be used to accelerate the inspiral calculation.
In Section~\ref{section:Two_Timescale} we present a two-timescale expansion of the equations of motion.
One advantage of this approach is that the adiabatic and post-adiabatic inspirals can be computed offline.
Given a particular mass ratio, the true inspiral can then be computed from interpolating functions as a very fast online step.
It may be advantageous to design data analysis pipelines to leverage this additional speed-up.

We conclude this section by presenting averaged equations of motion for the case of quasi-spherical Kerr inspirals in Section~\ref{section:Averaged_EOM}.

\subsection{Review of near-identity transformations for generic EMRI systems} \label{section:NIT_Review}

The NIT variables, $\nit{P}_j$, $\nit{q}_i$ and $\nit{S}_k$, are related to the OG variables  $P_j$, $q_i$ and $S_k$ via
\begin{subequations}\label{eq:transformation}
	\begin{align}
	\begin{split}\label{eq:transformation1}
		\nit{P}_j &= P_j + \sp Y_j^{(1)}(\vec{P},\vec{q}) + \sp^2 Y_j^{(2)}(\vec{P},\vec{q}) + \HOT{3},
	\end{split}\\
	\begin{split}
		\nit{q}_i &= q_i + \sp X_i^{(1)}(\vec{P},\vec{q}) +\sp^2 X_i^{(2)}(\vec{P},\vec{q}) + \HOT{3},
	\end{split}\\
	\begin{split}
		\nit{S}_k &= S_k + Z_k^{(0)}(\vec{P},\vec{q}) +\sp Z_k^{(1)}(\vec{P},\vec{q}) + \HOT{2}.
	\end{split}
	\end{align}
\end{subequations}
Here, the transformation functions $Y_j^{(n)}$, $X_i^{(n)}$, and $Z_k^{(n)}$ are required to be smooth, periodic functions of the orbital phases $\vec{q}$. 
At leading order, Eqs.~\eqref{eq:transformation} are identity transformations for $P_k$ and $q_i$ but not for $S_k$ due to the presence of a zeroth order transformation term $Z_k^{(0)}$.

The inverse transformations can be found for $P_k$ and $q_i$ by requiring that their composition with the transformations in Eqs.~\eqref{eq:transformation} must give the identity transformation. To first order in $\epsilon$, this gives us
\begin{subequations}\label{eq:inverse_trasformaiton}
	\begin{align}
			\begin{split}
					P_j &= \nit{P_j} - \epsilon Y_j^{(1)}(\vec{\nit{P}},\vec{\nit{q}}) + \HOT{2},
				\end{split}\\
		\begin{split}
					q_i &= \nit{q_i} - \epsilon X_i^{(1)}(\vec{\nit{P}},\vec{\nit{q}}) +  \HOT{2}.
				\end{split}
		\end{align}
\end{subequations}
It will be useful to decompose various functions into Fourier series where we use the convention:
\begin{equation} \label{eq:Fourier}
	A(\vec{P},\vec{q}) = \sum_{\vec{\kappa} \in \mathbb{Z}^N} A_{\vec{\kappa}}(\vec{P}) e^{i \vec{\kappa} \cdot \vec{q}},
\end{equation}
where $N$ is the number of orbital phases.
Based on this, we can split the function into an averaged piece $\avg{A} (\vec{P})$ and an oscillating piece
\begin{equation}\label{eq:oscillating}
	\osc{A}(\vec{P},\vec{q}) = A(\vec{P},\vec{q}) - \avg{A}(\vec{P})  = \sum_{\vec{\kappa} \neq \vec{0}} A_{\vec{\kappa}}(\vec{P}) e^{i \vec{\kappa} \cdot \vec{q}}. 
\end{equation}

Using the above transformations along with the equations of motion, and working order by order in $\mr$, we can choose values for the transformation functions $Y_j^{(1)}, Y_j^{(2)}, X_i^{(1)}, X_i^{(2)}, Z_k^{(0)}$ and $Z_k^{(1)}$ such that the resulting equations of motion for $\nit{P_j}, \nit{q_i}$ and $\nit{S_k}$ take the following form
\begin{subequations}\label{eq:transformed_EoM}
	\begin{align}
	\begin{split}
		\frac{d \nit{P}_j}{d \lambda} &=\epsilon \nit{F}_j^{(1)}(\vec{\nit{P}}) + \epsilon^2 \nit{F}_j^{(2)}(\vec{\nit{P}}) +  \mathcal{O}(\epsilon^3),
	\end{split}\\
	\begin{split}
		\frac{d \nit{q}_i}{d\lambda} &= \Upsilon_i^{(0)}(\vec{\nit{P}}) +\epsilon \Upsilon_i^{(1)}(\vec{\nit{P}}) + \mathcal{O}(\epsilon^2),
	\end{split}\\
	\begin{split}
		\frac{d \nit{S}_k}{d \lambda} &= \Upsilon_k^{(0)}(\vec{\nit{P}}) + \epsilon \Upsilon_k^{(1)}(\vec{\nit{P}}) +  \mathcal{O}(\epsilon^2).
	\end{split}
	\end{align}
\end{subequations}
Crucially, these equations of motion are now independent of the orbital phases $\vec{q}$. 
Deriving the relationship between the transformed forcing functions, $\nit{F}_j^{(1)},\nit{F}_j^{(2)},\Upsilon_\alpha^{(0)} $ and $\Upsilon_\alpha^{(1)}$), and the original forcing functions is quite an involved process with several freedoms and choices, each with its own merits and drawbacks. 
This is discussed at length in paper I, so for brevity we will summarize the results and the particular choices we have made in this work.
The transformed forcing functions are related to the original functions by 
\begin{subequations}
	\begin{gather}
		\nit{F}_j^{(1)} = \left<F_{j}^{(1)}\right>, \quad \Upsilon_{i}^{(1)}= \left<f_{i}^{(1)}\right>, \quad \Upsilon_{k}^{(0)} = \left<f_{k} ^{(0)}\right>, \tag{\theequation a-c}
	\end{gather}
	\begin{equation}
		\nit{F}_j^{(2)} = \left<F_j^{(2)} \right> + \left<\frac{\partial \osc{Y}_j^{(1)}}{\partial \nit{q}_i} \osc{f}_i^{(1)} \right> + \left<\frac{\partial \osc{Y}_j^{(1)}}{\partial \nit{P}_k} \osc{F}_k^{(1)} \right>, \tag{\theequation d}
	\end{equation}
	\begin{equation}
		\Upsilon_k^{(1)} = - \left<\frac{\partial \osc{f}_k^{(0)}}{\partial \nit{P}_j} \osc{Y}_j^{(1)} \right> - \left<\frac{\partial \osc{f}_k^{(0)}}{\partial \nit{q}_i} \osc{X}_i^{(1)} \right>. \tag{\theequation e}
	\end{equation}
\end{subequations}
Note that since we currently lack second order in mass ratio contributions, we will set $\left<F_j^{(2)} \right> = 0$.
In deriving these equations of motion, we have constrained the oscillating pieces of the NIT transformation functions to be
\begin{equation}\label{eq:NIT_Y}
	\osc{Y}_j^{(1)} \equiv \sum_{\vec{\kappa} \neq \vec{0}} \frac{i}{\vec{\kappa} \cdot \vec{\Upsilon}^{(0)}} F_{j,\vec{\kappa}}^{(1)} e^{i \vec{\kappa} \cdot \vec{q}},
\end{equation}

\begin{equation}\label{eq:NIT_X}
	\osc{X}_i^{(1)} \equiv \sum_{\vec{\kappa} \neq \vec{0}}\left( \frac{i}{\vec{\kappa} \cdot \vec{\Upsilon}^{(0)}} f_{i,\vec{\kappa}}^{(1)} + \frac{1}{(\vec{\kappa} \cdot \vec{\Upsilon}^{(0)})^2} \frac{\partial \Upsilon_i^{(0)}}{\partial P_j}F_{j,\vec{\kappa}}^{(1)} \right) e^{i \vec{\kappa} \cdot \vec{q}},
\end{equation}
and $\osc{Z}_k^{(0)}$ is found by solving
\begin{equation}\label{eq:Z_definition}
	\osc{f}_k^{(0)} + \frac{\partial \osc{Z}_k^{(0)}}{\partial \nit{q}_i} \Upsilon_i^{(0)} = 0.
\end{equation}
This equation is satisfied by the oscillating pieces for the analytic solutions for the geodesic motion of $t$ and $\phi$,
\begin{equation}\label{eq:Z_solution}
	\osc{Z}_k^{(0)} = - \osc{S}_{k,r} (q_r) - \osc{S}_{k,z} (q_z).
\end{equation}
We have chosen the averaged pieces $\bavg{Y_j^{(1)}} = \bavg{Y_j^{(2)}}=\bavg{X_i^{(1)}} = \bavg{X_i^{(2)}} =\bavg{Z_k^{(0)}} =\bavg{Z_k^{(1)}} = 0$ for simplicity, though one could make other choices like in Ref.~\cite{Pound2021}.
In order to generate waveforms, one only needs to know the transformations in Eq.~\eqref{eq:transformation} to zeroth order in the mass ratio, i.e.,
\begin{subequations}
	\begin{align}
	\begin{split}
		P_j &= \nit{P_j} + \mathcal{O}(\epsilon),
	\end{split}\\
	\begin{split}
		q_i &= \nit{q_i} + \mathcal{O}(\epsilon),
	\end{split}\\
	\begin{split} \label{eq:Extrinsic_Transformation}
		S_k &= \nit{S}_k -Z_k^{(0)}(\vec{\nit{P}},\vec{\nit{q}} ) +  \mathcal{O}(\epsilon).
	\end{split}
	\end{align}
\end{subequations}
Furthermore, to be able to directly compare between OG and NIT inspirals, we will need to match their initial conditions to sufficient accuracy. To maintain an overall phase difference of $\mathcal{O}(\epsilon)$ throughout an inspiral, this requires the transformation of the $P_j$'s \eqref{eq:transformation1} to linear order in $\mr$, while it is sufficient to know the rest of Eq.~\eqref{eq:transformation} to zeroth order.

\subsection{Averaging transformations for motion parametrized by Boyer-Lindquist coordinate time} \label{section:t_NIT}
Solving the above equations results in solutions for $\nit{\vec{P}}$, $\nit{\vec{q}}$ and $\nit{\vec{S}}$ as functions of Mino time $\lambda$.
While this would include $t(\lambda)$, the transformation to $\lambda(t)$ is non-trivial, and in practice it is done via interpolation which can be costly for long inspirals. 
It would be significantly more convenient for the solutions to be functions of $t$ from the start so that one can produce waveforms for data analysis without this postprocessing step. 
This can be accomplished for the OG equations by simply using the chain rule:
\begin{subequations}
	\begin{align}
	\begin{split}
		\frac{d P_j}{d t} &= \frac{1}{f_t^{(0)}(\vec{P},\vec{q})} \left( \mr F_j^{(1)}(\vec{P}) \right),
	\end{split}\\
	\begin{split}
		\frac{d q_i}{d t} &= \frac{1}{f_t^{(0)}(\vec{P},\vec{q})} \left( \Upsilon_i(\vec{P}) + \mr f_i(\vec{P},\vec{q}) \right),
	\end{split}\\
	\begin{split}
		\frac{d \phi}{d t} &= \frac{1}{f_t^{(0)}(\vec{P},\vec{q})} \left( f_\phi^{(0)}(\vec{P},\vec{q}) \right).
	\end{split}
	\end{align}
\end{subequations}
Notice that we have one less equation of motion to solve.
However, using the same approach to the NIT equations of motion results in 
\begin{subequations}
	\begin{align}
	\begin{split}
		\frac{d {\nit{P}}_j}{dt} &= \frac{1}{f_t^{(0)}(\vec{P},\vec{q})} \left( \epsilon \nit{F}_j^{(1)}(\vec{\nit{P}}) + \mr^2 \nit{F}_j^{(2)}(\vec{\nit{P}}) \right),
	\end{split}\\
	\begin{split}
		\frac{d \nit{q}_i}{dt} &= \frac{1}{f_t^{(0)}(\vec{P},\vec{q})} \left( \Upsilon_i(\vec{\nit{P}}) +\mr \Upsilon_i^{(1)}(\vec{\nit{P}}) \right),
	\end{split}\\
	\begin{split}
		\frac{d\nit{\phi}}{dt} &=\frac{1}{f_t^{(0)}(\vec{P},\vec{q})} \left( \Upsilon_\phi(\vec{\nit{P}}) + \mr \Upsilon_\phi^{(1)}(\vec{\nit{P}}) \right).
	\end{split}
	\end{align}
\end{subequations}
As we can see, we have now reintroduced a dependence on the orbital phases $\vec{q}$, defeating the purpose of our original NIT.
Thankfully, as outlined in Ref.~\cite{Pound2021}, these oscillations can also be averaged out by performing another transformation:
\begin{subequations}\label{eq:t_param_transformation}
	\begin{align}
	\begin{split}\label{eq:transformation2}
		\mathcal{P}_j &= \nit{P}_j + \sp \Pi_j^{(1)}(\vec{\nit{P}},\vec{\nit{q}}) + \sp^2 \Pi_j^{(2)}(\vec{\nit{P}},\vec{\nit{q}}) + \HOT{3},
	\end{split}\\
	\begin{split}
		\varphi_\alpha &= \nit{Q}_\alpha + \Delta \varphi_\alpha +  \sp \Phi_\alpha^{(1)}(\vec{\nit{P}},\vec{\nit{q}}) + \HOT{2},
	\end{split}
	\end{align}
\end{subequations}
where $\vec{Q} = \{\vec{\nit{q}},\nit{\phi}\}$, $\Delta \varphi_\alpha= \Omega_\alpha^{(0)}(\vec{\nit{P}}) \Delta t^{(0)}$ and $\Omega_\alpha^{(0)}$ is the Boyer-Lindquist fundamental frequency of the tangent geodesic.

To obtain the equations of motion for $\mathcal{P}_j$ and $\varphi_i$, we take the time derivative of Eq.~\eqref{eq:t_param_transformation}, substitute in the expression for the NIT equations of motion, and then use the inverse transformation of Eq.~\eqref{eq:t_param_transformation} to ensure that all functions are expressed in terms of $\vec{\mathcal{P}}$ and $\vec{\nit{q}}$ and expanding order by order in $\epsilon$. 
We then choose the oscillatory functions $\Delta t$, $\Phi_i^{(1)}$, $\Pi_j^{(1)}$ and $\Pi_j^{(2)}$ to cancel out any oscillatory terms that appear at each order in $\mr$. 

This results in averaged equations of motion that take the following form: 
\begin{subequations}\label{eq:t_transformed_EoM}
	\begin{align}
	\begin{split}
		\frac{d \mathcal{P}_j}{dt} &= \epsilon \Gamma_j^{(1)}(\vec{\mathcal{P}}) + \epsilon^2 \Gamma_j^{(2)}(\vec{\mathcal{P}}) +  \mathcal{O}(\epsilon^3),
	\end{split}\\
	\begin{split} \label{eq:Phase_Transformation}
		\frac{d\varphi_\alpha}{dt} &= \Omega^{(0)}_\alpha(\vec{\mathcal{P}}) +\epsilon \Omega_\alpha^{(1)}(\vec{\mathcal{P}}) + \mathcal{O}(\epsilon^2).
	\end{split}
	\end{align}
\end{subequations}
These equations of motion are related to the Mino time averaged equations of motion, Eqs.~\eqref{eq:transformed_EoM}, where the adiabatic terms are given by
\begin{subequations}
	\begin{align}
		\Gamma^{(1)}_j = \frac{\nit{F}_j^{(0)}}{\Upsilon_t^{(0)}}, \quad \Omega^{(0)}_\alpha = \frac{\Upsilon_\alpha^{(0)}}{\Upsilon_t^{(0)}},
		\tag{\theequation a-b}
	\end{align}
\end{subequations}
and the post-adiabatic terms are given by
\begin{subequations}
	\begin{align}
		\begin{split}
		\Gamma^{(2)}_j & = \frac{1}{\Upsilon_t} \Big( 
		\nit{F}^{(2)}_j + \nit{F}^{(1)} \frac{\partial}{\partial \mathcal{P}_j} \avg{\Pi^{(1)}_j} \\
		& - \avg{f_t^{(0)} \Pi_k^{(1)}} \PD{\Gamma_j^{(1)}}{\mathcal{P}_k} - \Upsilon_t^{(1)} \Gamma^{(1)}_j \Big),
		\end{split}\\
		\begin{split}
			\Omega^{(1)}_\alpha & = \frac{1}{\Upsilon_t^{(0)}} \Big( \Upsilon^{(1)}_\alpha + \nit{F}^{(1)}_j \avg{\PD {\Delta \varphi_\alpha}{\mathcal{P}_j}} \\
			& - \avg{f_t^{(0)} \Pi_k^{(1)}} \PD{\Omega_\alpha^{(0)}}{\mathcal{P}_k}  - \Upsilon_t^{(1)} \Omega^{(1)}_i \Big).
		\end{split}
	\end{align}
\end{subequations}
This constrains the oscillating pieces of our transformation to be 
\begin{subequations}
	\begin{align}
	\begin{split} \label{eq:Delta_t}
		\Delta t = &\sum_{\kappa \neq 0} \frac{f_{t,\vec{\kappa}}^{(0)}}{ -i \vec{\kappa} \cdot \vec{\Upsilon}^{(0)}} = -\osc{Z}_t,
	\end{split}\\
	\begin{split} \label{eq:leading_order_t_NIT}
		\osc{\Pi}^{(1)}_j = &\sum_{\kappa \neq 0} \frac{f_{t,\vec{\kappa}}^{(0)}}{ -i \vec{\kappa} \cdot \vec{\Upsilon}^{(0)}} \Gamma_j^{(1)} = -  \osc{Z}_t \Gamma_j^{(1)},
	\end{split}\\
	\begin{split}
		\Phi^{(1)}_{\alpha,\vec{\kappa}} = & \frac{i}{\vec{\kappa} \cdot \vec{\Upsilon}^{(0)} }\Biggl( 
		\PD{\Delta \varphi_{\alpha,\vec{\kappa}}}{\mathcal{P}_j} \nit{F}_j^{(1)}  - \frac{f^{(0)}_{t,\vec{\kappa}}}{\Upsilon^{(0)}_t} \Upsilon^{(1)}_t \Omega^{(0)}_t
		\\& +\sum _{\vec{\kappa}' \neq \vec{0}} \Biggl[ \left( i \vec{\kappa}' \cdot \vec{X}^{(1)}_{\vec{\kappa} - \vec{\kappa}'} f_{t,\vec{\kappa}'}^{(0)} + Y^{(1)}_{j,\vec{\kappa} - \vec{\kappa}'} \PD{f^{(0)}_{t,\vec{\kappa}'}}{\mathcal{P}_j} \right) \Omega^{(0)}_\alpha 
		\\& - \Pi^{(1)}_{j,\vec{\kappa} - \vec{\kappa}'} f_{t, \vec{\kappa}'}^{(0)} \PD{\Omega^{(0)}_\alpha}{\mathcal{P}_j} \Biggr] \Biggr).
	\end{split}
	\end{align}
\end{subequations}

We are free to chose the averaged pieces of $\Pi_j^{(1)}$, and we make the simplification that $\avg{\Pi_j^{(1)}} = 0$. 
With this and the identity $\avg{ f_t^{(0)} (\int{f}_t^{(0)} d \vec{q} )} = 0 $, we get the further simplification $\avg{f_t^{(0)} \Pi_j^{(1)}} = 0$. 
Thus, the expressions for $\Gamma^{(2)}_j$ and $\Omega_\alpha^{(1)}$ simplify to
\begin{subequations}
	\begin{align}
	\begin{split}
		\Gamma^{(2)}_j = \frac{1}{\Upsilon_t^{(0)}} \Big( 
		\nit{F}^{(2)}_j -\Upsilon_t^{(1)} \Gamma^{(1)}_j \Big),
	\end{split}\\
	\begin{split}
			\Omega^{(1)}_\alpha = \frac{1}{\Upsilon_t^{(0)}} \Big( \Upsilon^{(1)}_\alpha -\Upsilon_t^{(1)}\Omega^{(0)}_\alpha \Big).
	\end{split}
	\end{align}
\end{subequations}

What is most useful about these equations of motion is that their solutions $\vec{\mathcal{P}}(t)$ and $\vec{\varphi}(t)$ is exactly what is required to feed into waveform generating schemes. 
We show the equivalence between $\vec{\varphi}(t)$ and the relationship derived in Ref.~\cite{McCart2021} between the solutions to the original NIT equations of motion and the waveform phases $\Phi_{m n}$ in Appendix~\ref{section:tNIT_and_Wavefrom_Phases}.

\subsection{Two-timescale expansion} \label{section:Two_Timescale}
There is a related way of obtaining the above solutions via the TTE.
We exploit the difference between the timescales of the system by defining $\slowt \coloneqq \mr t$ as the slow time which governs the long-term, secular behaviour of the system and defining $t$ as the fast time of the system that governs the short-term, orbital dynamics and treating these two times independent variables. 

As such, we expand the transformed variables as 
\begin{subequations} \label{eq:two-timescale-expansion}
	\begin{align}
	\begin{split}
		\mathcal{P}_j(\slowt,\mr) = \mathcal{P}^{(0)}_j(\slowt) + \mr \mathcal{P}^{(1)}_j(\slowt) + \HOT{2},
	\end{split}\\
	\begin{split} 
		\varphi_\alpha(\slowt,\mr) = \frac{1}{\mr} \left[\varphi^{(0)}_\alpha(\slowt) + \mr \varphi^{(1)}_\alpha(\slowt) \right] + \HOT{1}.
	\end{split}
	\end{align}
\end{subequations}

Applying this expansion to the $t$ parametrized NIT equations of motion, one finds that the equations of motion for the two-timescale expanded variables takes the form:
\begin{subequations}
	\begin{align}
	\begin{split}
		\frac{d \mathcal{P}^{(0)}_j }{d \slowt} &= \Gamma^{(0)}_j(\vec{\mathcal{P}}^{(0)}),
	\end{split}\\
	\begin{split}
		\frac{d \varphi^{(0)}_\alpha }{d \slowt} &= \Omega^{(0)}_\alpha(\vec{\mathcal{P}}^{(0)}),
	\end{split}\\
	\begin{split}
		\frac{d \mathcal{P}^{(1)}_j }{d \slowt} &= \Gamma^{(2)}_j(\vec{\mathcal{P}}^{(0)}) + \mathcal{P}^{(1)}_k \left(\PD{\Gamma^{(1)}_j}{\mathcal{P}_k}(\vec{\mathcal{P}}^{(0)})\right),
	\end{split}\\
	\begin{split}
		\frac{d \varphi^{(1)}_\alpha }{d \slowt} &= \Omega^{(1)}_\alpha(\vec{\mathcal{P}}^{(0)}) + \mathcal{P}^{(1)}_k \left(\PD{\Omega^{(0)}_\alpha}{\mathcal{P}_k}(\vec{\mathcal{P}}^{(0)})\right).
	\end{split}
	\end{align}
\end{subequations}

There is a trade-off for solving these equations of motion. 
We now have to solve a system of coupled differential equations that is twice the size and thus is more expensive to solve numerically, but the solutions are independent of $\mr$ and one can construct a solution for any given value of $\mr$ using Eqs.~\refeq{eq:two-timescale-expansion}. 
Thus, if one wants to compute multiple inspirals with varying mass ratios, the TTE can be more efficient overall.
However, there is also an issue where the inspiral will stop earlier, as the variables $\mathcal{P}^{(0)}_j$ typically reach values which correspond to the ISSO before $\mathcal{P}_j$ do.
In this regime, one should instead employ a transition to plunge as this is where the adiabaticity assumptions of the OG equations and the two-timescale expansion is expected to break down. 
As such we will use both the NIT and the TTE equations of motion to produce waveforms and compare them to waveforms generated using the OG equations to assess which is the more practical framework for producing post-adiabatic EMRI waveforms.

\subsection{Averaged equations of motion for quasi-spherical Kerr inspirals} \label{section:Averaged_EOM}
We now specialize the above results to the case of quasi-spherical inspirals into a Kerr black hole.
The NIT equations of motion parameterized by Mino time take the form:
\begin{subequations}\label{eq:NIT_EoM}
	\begin{align}
		\begin{split}
			\frac{d \nitr}{d \lambda} &= \mr \nit{F}_r^{(1)} (a, \nitr, \nit{x}) + \mr^2 \nit{F}_r^{(2)} (a, \nitr, \nit{x}),
		\end{split}\\
		\begin{split}
			\frac{d \nit{x}}{d \lambda} &= \mr \nit{F}_x^{(1)} (a, \nitr, \nit{x}) + \mr^2 \nit{F}_x^{(2)} (a, \nitr, \nit{x}),
		\end{split}\\
		\begin{split}
			\frac{d \nit{q_z}}{d \lambda} &=  \Upsilon_z^{(0)}(a, \nitr,\nit{x}) + \mr \Upsilon_z^{(1)} (a, \nitr \nit{x}),
		\end{split}\\
		\begin{split}
			\frac{d \nit{t}}{d \lambda} &= \Upsilon_t^{(0)}(a, \nitr, \nit{x}) + \mr \Upsilon_t^{(1)}(a, \nitr, \nit{x}), 
		\end{split}\\
		\begin{split}
			\frac{d \nit{\phi}}{d \lambda} &= \Upsilon_{\phi}^{(0)}(a, \nitr, \nit{x}) + \mr \Upsilon_{\phi}^{(1)}(a, \nitr, \nit{x}).
		\end{split}
	\end{align}
\end{subequations}
The leading order terms in each equation of motion are simply the original functions, derived in Appendix~\ref{section:OG_Spherical} and paper II,  averaged over a single geodesic orbit, i.e.,
\begin{subequations}
\begin{equation} \label{eq:NIT_Relationship1}
	\nit{F}_r^{(1)} = \avg{F^{(1)}_{r}}, \quad \nit{F}_x^{(1)} = \avg{F^{(1)}_{x}}, \quad \Upsilon_z^{(1)}= \avg{f^{(1)}_{z}}, \tag{\theequation a-c}
\end{equation}
\begin{equation}\label{eq:NIT_Relationship2}
	\Upsilon_t^{(0)} =  \avg{f_{t}^{(0)}}  \quad \Upsilon_\phi^{(0)} =  \avg{f_{\phi}^{(0)}} \tag{\theequation d-e}, 
\end{equation}
\end{subequations}
where $\Upsilon_t^{(0)}$ and $\Upsilon_\phi^{(0)}$ are the Mino time $t$ and $\phi$ fundamental frequencies which are known analytically~\cite{Fujita2009a}.
The remaining terms are more complicated and require Fourier decomposing the original functions and their derivatives with respect to the orbital elements $(\r,x)$.
To express the result, for any function 
\begin{equation}
A(a,\r,x,q_z) = \sum_{\kappa} A_\kappa(a,\r,x) {e^{i\kappa q_z}},
\end{equation}
we define the operator
\begin{equation} \label{eq:N_Operator}
	\begin{split}
	\mathcal{N}(A) &=  \sum_{\kappa \neq 0} \frac{-1}{\Upsilon_z^{(0)}} 
	\biggl[ A_{\kappa} f_{z,-\kappa}^{(1)} - 
	\frac{i}{\kappa} \bigg( \PD{A_{\kappa} }{\nitr} F_{r,-\kappa} +
	\PD{A_{\kappa} }{\nit{x} } F_{x,-\kappa} 
	\\ &-
	\frac{A_{\kappa}}{\Upsilon_z^{(0)}} \left( \PD{\Upsilon_z^{(0)}}{\nitr} F_{r,-\kappa} +
	\PD{\Upsilon_z^{(0)}}{\nit{x}} F_{x,-\kappa} \right) \bigg) \biggr].
	\end{split}
\end{equation}
With this in hand, the remaining terms in the equations of motion are found to be
\begin{equation}\label{eq:NIT_Relationship3}
	\begin{split}
	\nit{F}_r^{(2)} &=  \mathcal{N}(F_r^{(1)}), \quad \nit{F}_x^{(2)} =  \mathcal{N}(F_x^{(1)}), 
	\\ \Upsilon_t^{(1)} &=  \mathcal{N}(f_t^{(0)}), \quad \Upsilon_\phi^{(1)} =  \mathcal{N}(f_\phi^{(0)}).
	\end{split}
	\tag{\theequation a-d}
\end{equation}
Combining these results with Eqs.~\eqref{eq:t_transformed_EoM}, one can find the NIT equations of motion parametrized by Boyer-Lindquist time $t$ for the phases $\vec{\varphi} = \{\varphi_z,\varphi_\phi\}$ and orbital elements $\vec{\mathcal{P}} = \{ \tnitr, x_\varphi\}$ in the form
\begin{subequations}
	\begin{align}
		\begin{split}
			\frac{d \tnitr}{dt} &= \mr \Gamma_r^{(1)}(a,\tnitr,x_\varphi) + \mr^2 \Gamma_r^{(2)}(a,\tnitr,x_\varphi),
		\end{split}\\
		\begin{split}
			\frac{d x_\varphi}{dt} &= \mr \Gamma_x^{(1)}(a,\tnitr,x_\varphi) + \mr^2 \Gamma_x^{(2)}(a,\tnitr,x_\varphi),
		\end{split}\\
		\begin{split}
			\frac{d\varphi_z}{dt} &= \Omega^{(0)}_z(a,\tnitr, x_\varphi) +\mr \Omega_z^{(1)}(a,\tnitr,x_\varphi),
		\end{split}\\
		\begin{split}
			\frac{d\varphi_\phi}{dt} &= \Omega^{(0)}_\phi(a,\tnitr,x_\varphi) +\mr \Omega_\phi^{(1)}(a,\tnitr,x_\varphi).
		\end{split}
	\end{align}
\end{subequations}
The leading order terms in these equations are given by
\begin{subequations}
\begin{equation} \label{eq:t_NIT_Relationship1}
	\Gamma_r^{(1)} = \nit{F}_r^{(1)}/\Upsilon_t^{(0)}, \quad \Gamma_x^{(1)} = \nit{F}_x^{(1)}/\Upsilon_t^{(0)},\tag{\theequation a-b}
\end{equation}
\begin{equation}\label{eq:t_NIT_Relationship2}
	\Omega_z^{(0)} =  \Upsilon_z^{(0)} /\Upsilon_t^{(0)},\quad \Omega_\phi^{(0)} =  \Upsilon_\phi^{(0)} /\Upsilon_t^{(0)}. \tag{\theequation c-d}
\end{equation}
\end{subequations}
The sub-leading terms are given by
\begin{subequations}
	\begin{align}
		\begin{split}
			\Gamma^{(2)}_r = \frac{1}{\Upsilon_t^{(0)}} \Big( 
			\nit{F}^{(2)}_r -\Upsilon_t^{(1)} \Gamma^{(1)}_r \Big),
		\end{split}\\
		\begin{split}
			\Gamma^{(2)}_x = \frac{1}{\Upsilon_t^{(0)}} \Big( 
			\nit{F}^{(2)}_x -\Upsilon_t^{(1)} \Gamma^{(1)}_x \Big),
		\end{split}\\
		\begin{split}
			\Omega^{(1)}_z = \frac{1}{\Upsilon_t^{(0)}} \Big( \Upsilon^{(1)}_z -\Upsilon_t^{(1)}\Omega^{(0)}_z \Big),
		\end{split}\\
		\begin{split}
			\Omega^{(1)}_\phi = \frac{1}{\Upsilon_t^{(0)}} \Big( \Upsilon^{(1)}_\phi -\Upsilon_t^{(1)}\Omega^{(0)}_\phi \Big).
		\end{split}
	\end{align}
\end{subequations}

Finally, using the two-timescale expansion the adiabatic equations of motion are given by
\begin{subequations} \label{eq:Spherical_0PA_EoM}
		\begin{gather}
			\frac{d \tter{0}}{d\slowt} = \Gamma_r^{(1)}, \quad
			\frac{d x_\varphi^{(0)}}{d\slowt} = \Gamma_x^{(1)},
			\tag{\theequation a-b}
		\end{gather}
		\begin{gather}
			\frac{d \varphi_z^{(0)}}{d\slowt} = \Omega_z^{(0)}, \quad
			\frac{d \varphi_\phi^{(0)}}{d\slowt} = \Omega_\phi^{(0)}.
			\tag{\theequation c-d}
		\end{gather}
\end{subequations}
The post-adiabatic contributions to the equations of motion are given by 
\begin{subequations} \label{eq:Spherical_1PA_EoM}
	\begin{align}
		\begin{split}
			\frac{d \tter{1}}{d\slowt} &= \Gamma_r^{(2)} + \tter{1}\PD{\Gamma_r^{(1)}}{\tter{0}} + x_\varphi^{(1)} \PD{\Gamma_r^{(1)}}{x_\varphi^{(0)}},
		\end{split}\\
		\begin{split}
			\frac{d x_\varphi^{(1)}}{d\slowt} &= \Gamma_x^{(2)} + \tter{1} \PD{\Gamma_x^{(1)}}{\tter{0} } + x_\varphi^{(1)} \PD{\Gamma_x^{(1)}}{x_\varphi^{(0)}},
		\end{split}\\
		\begin{split}
			\frac{d \varphi_z^{(1)}}{d\slowt} &= \Omega_p^{(1)} + \tter{1}  \PD{\Omega_z^{(0)}}{\tter{0} } + x_\varphi^{(1)} \PD{\Omega_z^{(0)}}{x_\varphi^{(0)}},
		\end{split}\\
		\begin{split}
			\frac{d \varphi_\phi^{(1)}}{d\slowt} &= \Omega_\phi^{(1)} + \tter{1} \PD{\Omega_\phi^{(0)}}{\tter{0} } + x_\varphi^{(1)} \PD{\Omega_\phi^{(0)}}{x_\varphi^{(0)}},
		\end{split}
	\end{align}
\end{subequations}
Solutions for post-adiabatic inspirals can be obtained by solving Eqs.~\eqref{eq:Spherical_0PA_EoM} and \eqref{eq:Spherical_1PA_EoM} simultaneously and using Eqs.~\eqref{eq:two-timescale-expansion} along with a value for the mass ratio $\mr$ to recover $\tnitr (t),x_\varphi(t),\varphi_z(t)$ and $\varphi_z(t)$.

With the averaged equations for quasi-spherical Kerr inspirals in hand, we will now outline our numerical implementation for rapidly computing quasi-spherical self-forced inspirals.

\section{Implementation}\label{section:Implimentation}
Combining the interpolated GSF model along with our action-angle formulation of the OG equations provides us with all the information required to calculate the NIT and TTE equations of motion. 
We first evaluate and interpolate the various terms in the NIT and TTE equations of motion across the parameter space.
While this offline process can be expensive, it only needs to be completed once. 
Once completed, the online process of calculating self-forced inspirals can be completed in less than a second. This procedure is very similar to Papers I \& II, though now we also export interpolating functions for the partial derivatives needed for the TTE equations of motion.
\subsection{Offline Steps}

The offline calculation consists of the following steps. 
\begin{enumerate}
	
	\item  We start by selecting a grid which covers the parameter space. 
	We choose $y$ values between 0.099 and 0.999 in 451 equally spaced steps and $z_-^2$ values from 0.002 to 0.5 in 250 equally spaced steps (giving a total of 112,750 points).\footnote{
		Evaluating the NIT functions is computationally cheap so using a dense grid does not significantly increase the computational burden.
		Using an equally spaced grid also allows us to use \textit{Mathematica}'s default Hermite polynomial interpolation method for convenience of implementation.
		The grid spacing is chosen to be sufficiently dense so that the interpolation error is a negligible source of error for our comparisons between the OG, NIT and TTE inspirals, though a less dense grid may also achieve this.
	}

	\item For each point in the parameter space $(a,y,z_-^2)$ we evaluate the functions $F_{r \backslash x}^{(1)}$, $f_z^{(1)}$ and$f_{t \backslash \phi}^{(0)}$ along with their derivatives with respect to $r_p$ and $x$ for 49 equally spaced values of $q_z$ from $0$ to $2\pi$.
	
	\item We then perform a fast Fourier transform on the output data to obtain the Fourier coefficients of the forcing functions and their analytical derivatives with respect to $\r$ and $x$.
	
	\item With these, we then use Eqs.~\eqref{eq:NIT_Relationship1}, \eqref{eq:NIT_Relationship2}, \eqref{eq:N_Operator} and \eqref{eq:NIT_Relationship3} to construct $\nit{F}_{r\backslash x}^{(1\backslash2)}$, $\partial \nit{F}_{r\backslash x}^{(1)} / \partial \r$, $\partial \nit{F}_{r\backslash x}^{(1)} / \partial x$, and $\Upsilon_{t\backslash z \backslash \phi}^{(1)}$ at that point in parameter space. All other terms needed for the NIT and TTE can be derived from these terms or are already known analytically.
	
	\item We also use Eqs.~(\ref{eq:NIT_Y}) to construct the Fourier coefficients of the first-order transformation functions $Y_{r\backslash x}^{(1)}$. These are needed when comparing NIT and OG/TTE inspirals, to ensure that the initial conditions are comparable. Otherwise this step can be skipped.
	
	\item We then repeat this procedure across the parameter space for each point in our grid.
	
	\item Finally we interpolate the values for $\nit{F}_{r\backslash x}^{(1\backslash2)}$, $\partial \nit{F}_{r\backslash x}^{(1)} / \partial \r$, $\partial \nit{F}_{r\backslash x}^{(1)} / \partial p$, and $\Upsilon_{t\backslash z \backslash \phi}^{(1)}$ along with the coefficients of $Y_{r\backslash x}^{(1)}$ across this grid using Hermite interpolation and store the interpolants for future use. 
	
\end{enumerate}
We implemented the above algorithm in \textit{Mathematica} 12.2 and find the calculation takes about 5 hours to complete when parallelized across 20 CPU cores.
Since these offline steps need only be completed once, this is a comparatively small price to pay.

\subsection{Online Steps}

By contrast, the online steps are required for every inspiral calculation, but are computationally inexpensive. 
The online steps for computing a NIT or TTE inspiral are as follows.
\begin{enumerate}
	
	\item We load in the interpolants for $\nit{F}_{r\backslash x}^{(1\backslash2)}$ and $\Upsilon_{t\backslash z \backslash \phi}^{(1)}$, and define the NIT equations of motion. For TTE inspirals, one also needs to load in the interpolated derivatives $\partial \nit{F}_{r\backslash x}^{(1)} / \partial \r$ and $\partial \nit{F}_{r\backslash x}^{(1)} / \partial x$ and then define the TTE equations of motion.
	
	\item In order to facilitate comparisons between OG, NIT, and TTE inspirals, we also load interpolants of the Fourier coefficients of $\breve{Y}_{r/x}^{(1)}$ and Eqs.~\eqref{eq:transformation}, \eqref{eq:t_param_transformation} and \eqref{eq:leading_order_t_NIT} to construct first order near-identity transformations.
	
	\item We state the initial conditions of the inspiral $(\r(0), x(0), q_z(0))$ and use the NIT to leading order in the mass ratio to transform these into initial conditions for the NIT/TTE equations of motion, i.e.,  $(\tnitr(0), x_\varphi(0), \varphi_z(0))$. 
	
	\item We then evolve the NIT or TTE equations of motion using an ODE solver (in our case Mathematica's \texttt{NDSolve}). 
	
\end{enumerate}

As with the offline steps we implement the online steps in Mathematica. Note that steps (ii) and (iii) are only necessary because we want to make direct comparisons between NIT and OG inspirals with the same initial conditions. In general, the difference between the NIT and OG variables will always be $\mathcal{O}(\mr)$, and so performing the NIT transformation or inverse transformation to greater than zeroth order in mass ratio will not be necessary when producing waveforms to post adiabatic order, i.e. with phases accurate to $\mathcal{O}(\mr)$.

\subsection{Waveform Generation}
With a trajectory in hand, can now generate the gravitational waveform. 
Ideally, this would be done using interpolated Teukolsky fluxes as implemented by, e.g., the \texttt{FastEMRIWaveform} package for eccentric Schwarzschild inspirals~\cite{Chua2021a,Katz2021}. 
Unfortunately, such a model is not currently available for quasi-spherical Kerr inspirals, and so we generate all of our waveforms using the ``semi-relativistic" approximation used by the numerical kludge models~\cite{Babak2007}. 
In this approach the Boyer-Lindquist coordinates of the particle are mapped to flat spacetime coordinates which are fed into the quadrupole formula that is used to produce the final waveform strain $h = h_+ + i h_\times$.
This approximation fairs surprisingly well compared to Teukolsky snapshot waveforms, even in the strong field regime.
What is most important for our purposes is that all waveforms are calculated using the same kludge formula so that any differences in the waveforms are due to differences in the calculations of the inspiral trajectory.
For our implementation, we sample every $t_{\text{step}} = 2M$ and calculate the waveform strain at each time step to produce our numerical waveforms. 

In order to calculate the fractional overlap $\mathcal{O}(h_1,h_2)$ between two time-domain waveform strains $h_1$ and $h_2$, we first define the noise weighted inner product of these two waveforms as
\begin{equation}
	\langle h_1| h_2\rangle= 2 \int_0^\infty \frac{\tilde{h}_1^*(f) \tilde{h}_2(f) + \tilde{h}_1(f) \tilde{h}_2^*(f)}{S_n(f)} df,
\end{equation}
where $\tilde{h}(f)$ denotes the Fourier transform of the time domain waveform $h(t)$ and $h^*$ is the complex conjugate of $h$.
We take the power spectral density (PSD) of the detector noise $S_n(f)$ to be a flat noise curve.

From this, one can calculate the fractional waveform overlap $\mathcal{O}$ via
\begin{equation}
	\mathcal{O}= \frac{\langle h_1| h_2\rangle}{\sqrt{\langle h_1| h_1\rangle \langle h_2| h_2\rangle }}. 
\end{equation}

This overlap ranges from 1, when the two waveforms are identical, to 0, when the waveforms are perfectly orthogonal.
When dealing with waveform overlaps close to 1, it is often useful to talk in terms of the  fractional waveform mismatch $\mathcal{M} = 1 - \mathcal{O}$. 
In practice, we make use of the $\texttt{WaveformMatch}$ function from the SimulationTools package to calculate waveform overlaps~\cite{SimulationTools}.

\section{Results} \label{section:Results}
\subsection{OG vs NIT and TTE inspirals}
To test the accuracy of our NIT and TTE implementations, we compare inspirals calculated using the OG equations of motion against inspirals calculated using the NIT or TTE equations of motion. We use a case study of a typical system EMRI with a primary of mass $M = 10^6 M_\odot$.
We chose the initial conditions for this inspiral to have an inclination of $x = 0.75$ and radius $\r = 7.75 M$. 
This highly inclined, strong field inspiral provides a good test of our numerical implementations. 
We see similar results for other initial conditions.
We also set the initial phases $q_z(0) = \phi(0) = 0$ for simplicity. 

\begin{figure}
	\includegraphics[width =\linewidth]{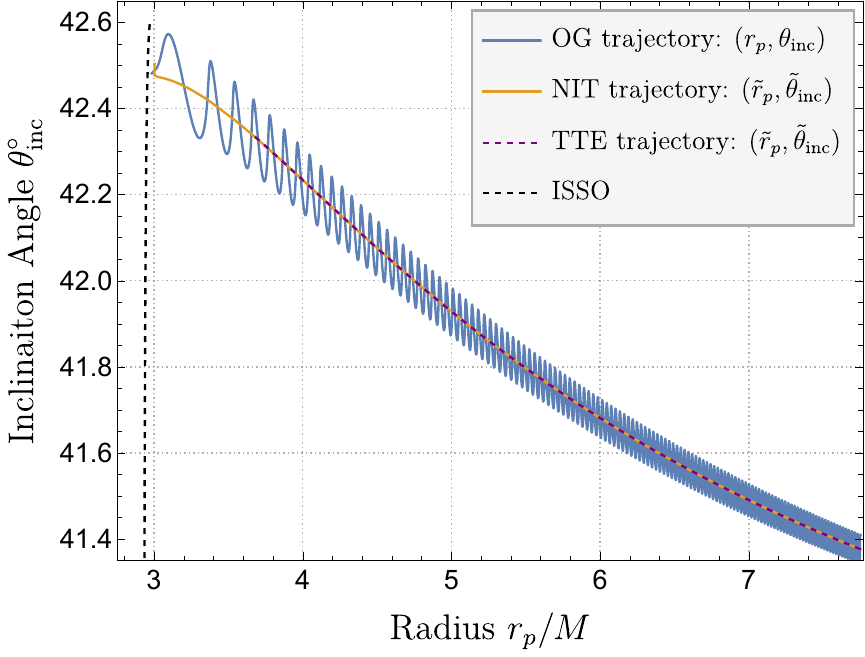}
	\caption{
	Trajectory through $(\r,\theta_\text{inc})$ space for an inspiral with $\mr =10^{-2}$, $a = 0.9 M$, and initial conditions $\r(0) = 7.75 M, x(0) = 0.75$. 
	We use such a large mass ratio to highlight the orbital timescale oscillations one encounters when using the OG equations. 
	Using the NIT or TTE equations of motion averages out these oscillations and results in almost identical inspirals. 
	We also see that the TTE equations of motion break down further from the ISSO than the OG or NIT equations of motion.}
	\label{fig:TrajectoryPlot}
\end{figure}

 Figure~\ref{fig:TrajectoryPlot} demonstrates the difference in the trajectories produced by each method.
We have purposely chosen an unusually large mass ratio for an EMRI of $\mr = 10^{-2}$ so that the orbital timescale oscillations are clearly visible  on the plot.
We see that for all trajectories the orbital separation decreases substantially, while the orbit becomes slightly more inclined.  
This is consistent with previous results for adiabatic quasi-spherical inspirals~\cite{Ryan:1995xi,Hughes2000,Hughes2001}.
The OG trajectory oscillates on the orbital timescale, while the NIT and TTE trajectories faithfully capture the average evolution of the trajectory. 
While the NIT and TTE trajectories may appear at first glance to be identical, it is important to note that the TTE equations of motion break down sooner than the NIT equations of motion.
This is due to the evolution of $r_\varphi^{(0)}$ and $x_\varphi^{(0)}$ reaching the location of the ISSO sooner than $r_\varphi$ and $x_\varphi$.
This is not as important of an issue as one might expect, as this close to the ISSO one should swap over to a ``transition to plunge" scheme~\cite{Thorne00,Burke2019,Compere21,Compere:2021zfj}.

\begin{figure}
	\includegraphics[width =\linewidth]{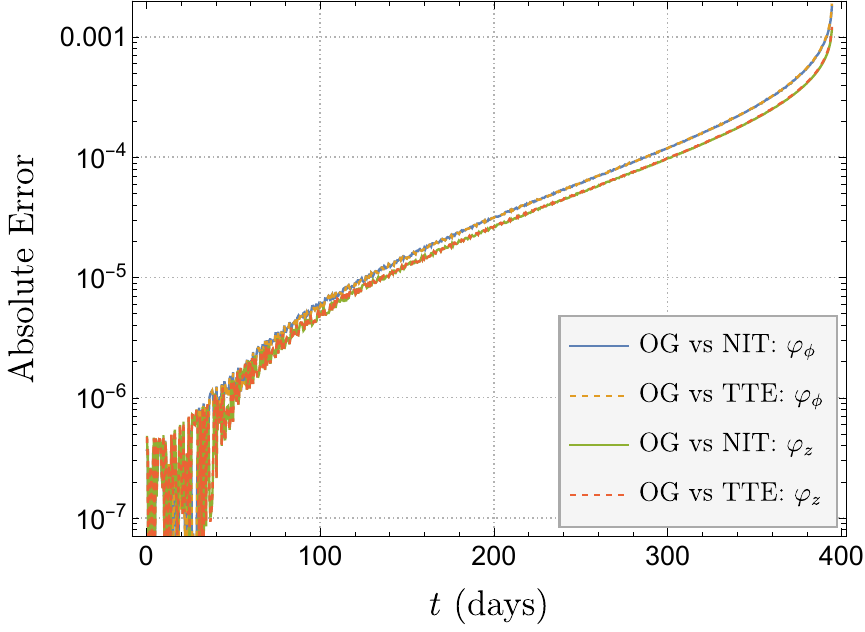}
	\caption{The difference in the orbital phases for a quasi-spherical Kerr inspiral with $M = 10^6 M_{\odot}$, $\mr = 10^{-5}$, and $a = 0.9M$ when using the OG equations versus the NIT or TTE equations of motion. In both cases, the difference stays small throughout the inspiral, only becoming large as the secondary approaches the ISSO when the adiabatic assumption breaks down. }
	\label{fig:PhasePlot}
\end{figure}
Figure~\ref{fig:PhasePlot} shows the difference in the phases between the OG trajectory and the NIT and TTE trajectories for a year long inspiral with a primary of mass $M = 10^6 M_\odot$ and mass ratio of $\mr = 10^{-5}$.
 In both cases, we find that the difference in the phases stays below $\sim10^{-3}$ throughout the inspiral, spiking only when the inspiral approaches the ISSO where the adiabatic assumption implicit in the OG, NIT and TTE equations of motion starts to break down. 
 Even then, the difference in the phases is substantially lower than subradian accuracy requirement needed for LISA data analysis.
 We also find that the growth in the error over time is most closely correlated with the interpolation error for the terms in the NIT and TTE equations of motion, and so interpolating on a denser grid should reduce this error even more.

\begin{figure}
	\includegraphics[width =\linewidth]{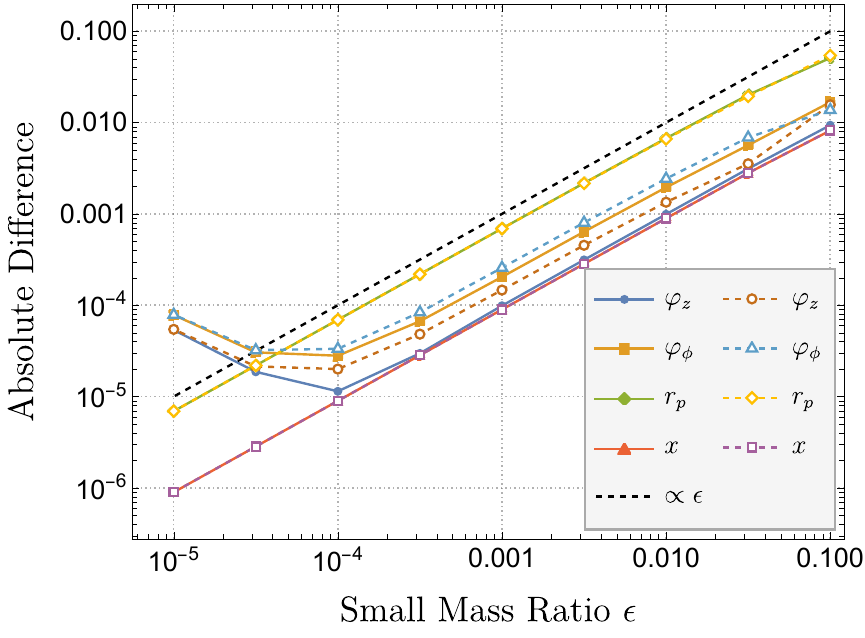}
	\caption{The absolute difference in the quantities of a prograde inspiral with $a = 0.9M$ and an initial inclination of $x_0 = 0.75$ after evolving from $r_p = 5 M$ to $r_p = 4 M$ when using different evolution equations. The solid lines show the difference between OG and NIT equations of motion and the dashed lines show the difference between using the OG and TTE equations of motion. We see that the differences generally follow the black $\mr$ curve and so converge linearly with the mass ratio until we reach mass ratios $\leq 10^{-3}$ where the error in the phases becomes dominated by interpolation and numerical error. }
	\label{fig:ErrorvsMassRatio}
\end{figure}

We note that formally both the NIT and the TTE should induce an error in both the phases and the orbital elements that scales linearly with the mass ratio.
To ensure that our implementation is converging correctly, we evolve inspirals from initial conditions $\r(0) = 5 M$ and $x(0) = 0.75$ until they reach $r_p = 4 M$ with different values of the mass ratio.
Since the OG quantities will have oscillations of $\mathcal{O}(\mr)$, sampling the error at only the final time step would make it difficult to determine the convergence with the mass ratio. As such we calculate the error at several time-steps in the last three orbital cycles and report the largest error.
The results of this test can be seen in Fig.~\ref{fig:ErrorvsMassRatio}. 
We see that for larger mass ratios the errors converge linearly, as expected. 
However, for mass ratios $ \lesssim 10^{-4}$, the formal error in the NIT and TTE is no longer the dominant source of error for the evolution of the phases. 
Instead the error is dominated by either interpolation error or the error in the ODE solver when solving the OG equations as we have found empirically that  this error is reduced by increasing the number of grid points used for interpolation and decreasing the relative tolerance of the integrator.
While both of these sources can be further suppressed at the cost of more expensive offline and online steps respectively, we show in Sec. \ref{section:Waveforms} that it is not necessary for producing trajectories accurate enough for LISA data science.

	\begin{table}
	\begin{center}
		\begin{tabular}{|c | c | c | c| c| c|} 
			\hline
			$\mr$ & OG  & NIT & Speedup & TTE   & Speedup \\[1ex] 
			\hline
			$10^{-2}$ & 24.7s & 0.48s & $\sim 51$  & 22ms & $\sim 1.1 \times 10^6$ \\ 
			%\hline
			$10^{-3}$ & 4m 20s & 0.17s & $\sim 1530$ & 39ms & $\sim 6.7 \times 10^6$ \\
			%\hline
			$10^{-4}$ & 43m 19s & 0.43s & $\sim 6044$ & 27ms & $\sim 9.6 \times 10^7$ \\
			%\hline
			$10^{-5}$ & 7hrs 42m & 0.49s & $\sim 54380$ & 22ms & $\sim 1.2 \times 10^9$ \\ [1ex] 
			\hline 
		\end{tabular}
	\end{center}
	\caption{Computational time required to evolve an inspiral from its initial conditions of $r_p(0) = 7.75 M$ and $x(0)= 0.75$ to the last stable orbit for different values of the mass ratio, as calculated in \textit{Mathematica} 13 on an Intel Core i7 @ 2.2GHz. The computational time for the OG inspiral scales inversely with the mass ratio, whereas the computational time for NIT inspirals is independent of the mass ratio. 
		The computation time for the TTE inspiral is $0.53s$, which is slightly longer than any of the NIT inspirals.
		However, if we consider the TTE calculation as an offline step, one can then immediately recover the solution for any value of $\mr$ in a matter of milliseconds.
		\label{table:Runtime}}
\end{table}

Now that we have established our averaging procedure does not change the accuracy of our model, we demonstrate the speedup that one enjoys from using either the NIT or TTE equations of motion instead of using the OG equations of motion in Table.~\ref{table:Runtime}. 
For each of these calculations, the initial conditions are set to $\r(0) = 7.75 M$ and $x(0) = 0.75$, and the inspirals are evolved to just before the ISSO, when $y(r_p, x) = 0.998$.
All calculations were done using machine precision arithmetic and an accuracy goal of $7$~digits for \textit{Mathematica}'s \texttt{NDSolve} function. 
We find that using the OG inspiral calculation takes longer as the mass ratio gets smaller since the solver will have to resolve many more orbital cycles before the inspiral reaches the ISSO.
However, the NIT inspirals take roughly the same amount of time regardless of the mass ratio.
The TTE inspiral takes slightly longer than the NIT inspiral, as one has to solve for twice as many equations. 
However once this step is complete, inspirals with the same initial conditions but any mass ratio can be computed extremely rapidly by evaluating the resulting interpolating functions.

\subsection{Waveform mismatches} \label{section:Waveforms}

 \begin{figure*}[!tb] \label{fig:WFPlot}
	\centering
	\subfloat[The first 3.5 hours of the waveform. \label{fig:WFEarly}]{
		\includegraphics[width=\textwidth]{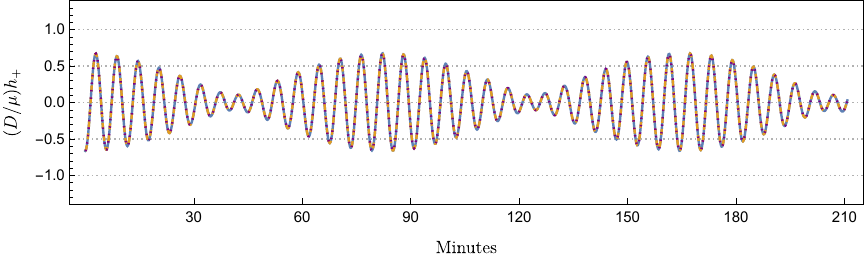}
	}
	\hfill
	\subfloat[The last 3.5 hours of the waveform. \label{fig:WFLate}]{
		\includegraphics[width=\textwidth]{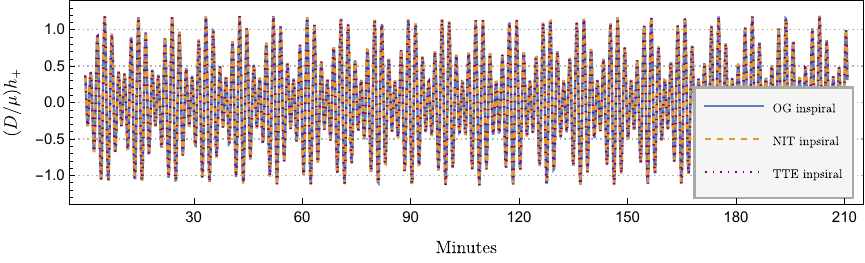}
	}
	\hfill
	\caption{``Semi-relativistic" quadrupole waveform strains for an EMRI with $M = 10^6 M_\odot$, $\mr = 10^{-5}$, $r_p(0) = 7.75M$, and $x(0) = 0.75$ as viewed edge on (i.e., with the detector at a latitude of $\Theta = \pi/2$ with respect to the source's frame), sampled once every 10s. The waveform strain is normalised by the luminosity distance from the source to the detector $D$ and the mass of the secondary $\mu$. Panel (a) shows the first 3.5 hours, and panel (b) shows the last 3.5 hours of this year-long waveform. The blue curve is the waveform generated from the OG inspiral, the dashed yellow curve is the waveform generated from the NIT inspiral, and the purple dotted curve is the waveform generated from the TTE inspiral. }
\end{figure*}
\mycomment{
	\begin{figure*} \label{fig:WFPlot}
		\centering
		\begin{subfigure}[b]{\textwidth}
			\centering
			\includegraphics[width=\textwidth]{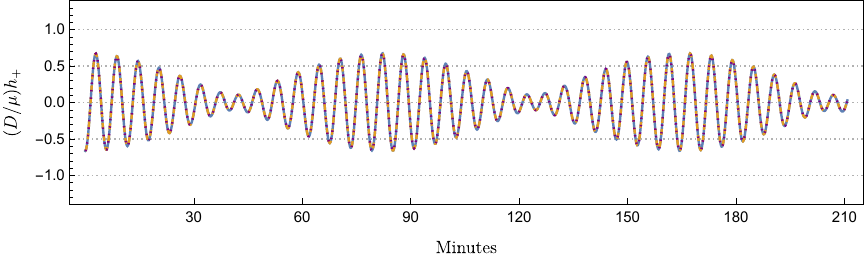}
			\caption{The first 3.5 hours of the waveform.}
			\label{fig:WFEarly}
		\end{subfigure}
		\hfill
		\begin{subfigure}[b]{\textwidth}
			\centering
			\includegraphics[width=\textwidth]{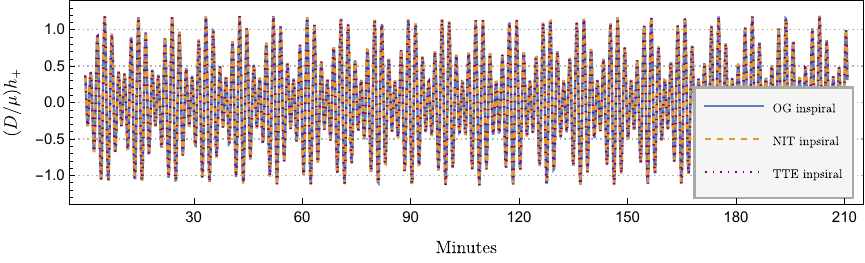}
			\caption{The last 3.5 hours of the waveform.}
			\label{fig:WFLate}
		\end{subfigure}
		\hfill
		\caption{``Semi-relativistic" quadrupole waveform strains for an ERMI with $\mr = 10^{-5}$, $p(0) = 7.75 M$ and $x(0) = 0.75$ as viewed from edge on sampled once every 10s. The blue curve is the waveform generated from the OG inspiral, the dashed yellow curve is the waveform generated from the NIT inspiral, and the purple dotted curve is the waveform generated from the TTE inspiral. }
	\end{figure*}
}
To generate waveforms, we use a semi-relativistic approximation to produce quadrupole waveforms which can be seen in Figs.~\ref{fig:WFEarly} and \ref{fig:WFLate}~\cite{Babak2007}.
These figures show the first and last 3.5 hours of the waveforms produced by the OG, NIT and TTE trajectories respectively, sampled once every 10s. 
The source is viewed edge on, i.e., the detector is located at a latitude $\Theta =  \pi/2$ and an azimuth of $\Phi = 0$ with respect to the source.
From the plots it is clear that the two waveforms overlap significantly.
The waveform mismatch between the OG waveforms and both the NIT and TTE waveforms is $\sim 2.5 \times 10^{-8}$.
This means that they meet the indistinguishability criteria~\cite{Flanagan1997, Lindblom2008, McWilliams2010} from the OG waveforms for signal-to-noise ratios (SNRs) of up to at least~$4500$.

\begin{figure} 
	\centering
	\subfloat[Mismatch between OG and NIT waveforms.\label{fig:MismatchNIT}]{
		\includegraphics[width=0.49\textwidth]{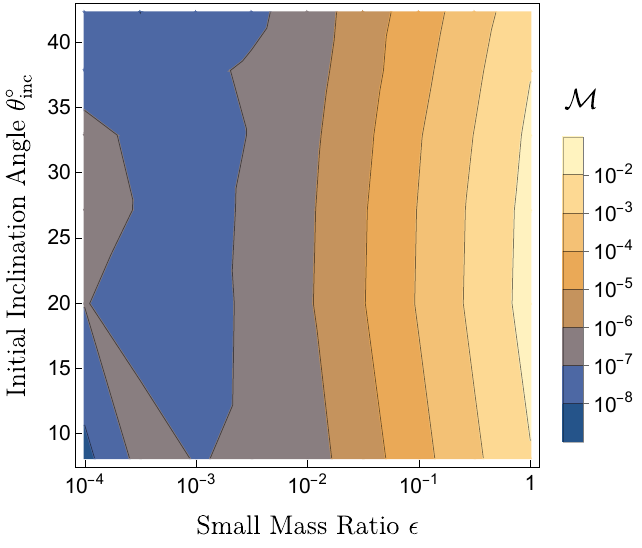}
	}
	\hfill
	\subfloat[Mismatch between OG and TTE waveforms. \label{fig:MismatchTTE}]{
		\includegraphics[width=0.49\textwidth]{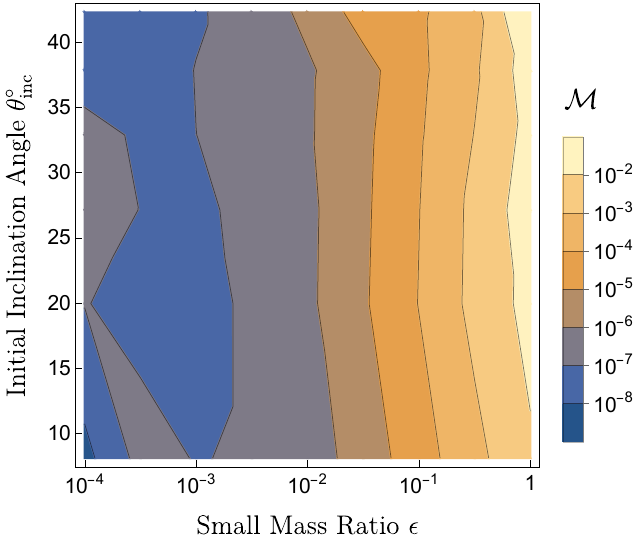}
	}
	\hfill
	\caption{Mismatch between year long OG and NIT/TTE waveforms as a function of orbital inclination and mass ratio. We do not see a substantial difference in accuracy between using either NIT or TTE waveforms. We also see that the mismatch remains smaller than 0.03 for mass ratios as large $\mr \approx 0.5$.\label{fig:MismatchPS} }
\end{figure}

Intuitively, one would expect the error in our averaging procedure to scale with the size of the orbital timescale oscillations which themselves scale with the mass ratio and the orbital inclination. 
As such, we wish determine the section of the parameter space where the difference in the waveforms between the OG and the NIT/TTE inspirals would be small enough not to matter for LISA data science. 
To do this, we first fix the mass of the primary to be $10^6 M_\odot$ and create a function which uses adiabatic inspirals and root finding to numerically compute the initial orbital separation for an inspiral which will take one year to reach the ISSO for a given inclination and mass ratio. 
We then create a grid of mass ratios $\mr = \{1, 10^{-1/2},10^{-1}, 10^{-3/2},10^{-2},10^{-5/2} ,10^{-3}, 10^{-7/2}, 10^{-4}\}$ and inclinations $x =  \{0.74,0.79,0.84,0.89,0.94,0.99\}$.
For each point on this grid, we calculate a year-long inspiral using the OG equations of motion, the NIT equations of motion, and the TTE equations of motion and then generate a waveform from each. 

The waveform mismatch between the OG and NIT or TTE waveforms is displayed in Fig.~\ref{fig:MismatchPS}. From these plots we see that there does not seem to be a substantial difference in terms of accuracy from using either the NIT or TTE equations of motion. 
While there may be some correlation between mismatch and inclination, this does not appear to be a strong effect for $\theta_{\text{inc}} \leq 45^\circ$.
The strongest effect on the waveform mismatch comes from the mass ratio. 
It is worth noting that $3 \times 10^{-2}$ is a commonly chosen maximum mismatch for a waveform template bank that corresponds to a 90 \%-ideal observed event rate~\cite{Owen1995}.
Our results suggest that one could, in principle, produce such a template bank of quasi-spherical inspirals using NIT or TTE waveforms even for mass ratios as large as $\mr \approx 0.5$.
In practice, we are still missing important post-adiabatic contributions such as second-order effects and contributions from the spin of the secondary; thus such a template bank would have substantial systematic biases. Also note that, formally, when all post-adiabatic corrections are included, the OG, NIT, and TTE inspirals have the same accuracy in terms of powers of the mass ratio. Consequently, we cannot say a priori which will be more faithful to Nature.

\subsection{Using Higher Precision Fluxes} \label{section:Flux_GSF_Hybrid}

\begin{figure}
	\includegraphics[width =\linewidth]{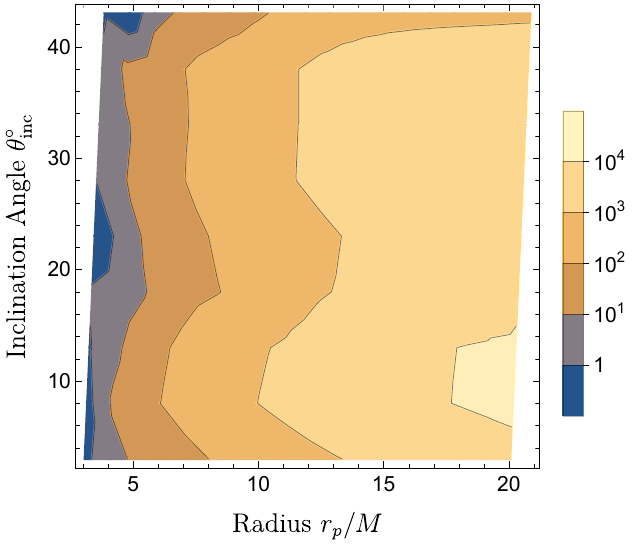}
	\caption{
	Absolute difference in $\varphi_\phi$ for a NIT inspiral evolved from a point in the parameter space to the ISSO with a typical EMRI mass ratio of $\mr = 10^{-5}$, depending on whether one incorporates high precision fluxes or not.
	 Using the high precision fluxes results in a more faithful inspiral, with phase differences with respect to the lower accuracy model in the range $\sim 10 - 10^4$ radians. 
	 As such, interpolating the adiabatic contributions to the averaged equations of motion to high precision is vitally important for accurate adiabatic and post-adiabatic EMRI waveforms.}
	\label{fig:AccurateFluxDifference}
\end{figure}
Now that we can compute fast and accurate inspirals, it is worth recalling that the relative accuracy of our interpolated GSF model is currently too low for production-level waveforms.
This is primarily due to the harsh relative accuracy requirements for the adiabatic pieces of $\lesssim \mr$ for subradian accuracy in the phases.
This can be improved by incorporating information from the asymptotic fluxes, which can be interpolated to a much higher accuracy across the parameter space due to the much cheaper cost of flux calculations compared to GSF calculations.
This is why our interpolated flux model is accurate to $\sim 10^{-6}$ while our interpolated GSF model is only accurate to $\sim 10^{-3}$.
Such information can be incorporated into the GSF model itself to improve the accuracy of the OG inspirals as well as the resulting NIT and TTE inspirals~\cite{Osburn2016}.
However since the  NIT and TTE equations of motion are naturally split into adiabatic and post-adiabatic pieces, one can calculate $\Gamma_j^{(1)}$ directly from the fluxes via Eqs.~\eqref{eq:flux_balance} and \eqref{eq:flux_chain_rule}, interpolate them to higher precision than using a GSF model, and then substitute these improved adiabatic terms into the averaged equations of motion.

To test the difference this would make to the overall accuracy of our post-adiabatic inspirals, we looked at the final error in the $\varphi_\phi$ phase when evolved from one point in the parameter space to the ISSO when using the flux+GSF model verses using only the GSF model for inspirals with $\mr = 10^{-5}$.
From Fig.~\ref{fig:AccurateFluxDifference}, we see that improving the adiabatic pieces results in a phase difference that can range for from tens to tens of thousands of radians for multi-year long inspirals and this difference gets larger as one moves away from the ISSO. 

As such, using high precision fluxes for the adiabatic contributions to the equations of motion is vital for obtaining accurate adiabatic and post-adiabatic inspirals.

\subsection{Impact of the self-force on spherical inspirals}

\begin{figure}
	\includegraphics[width =0.9\linewidth]{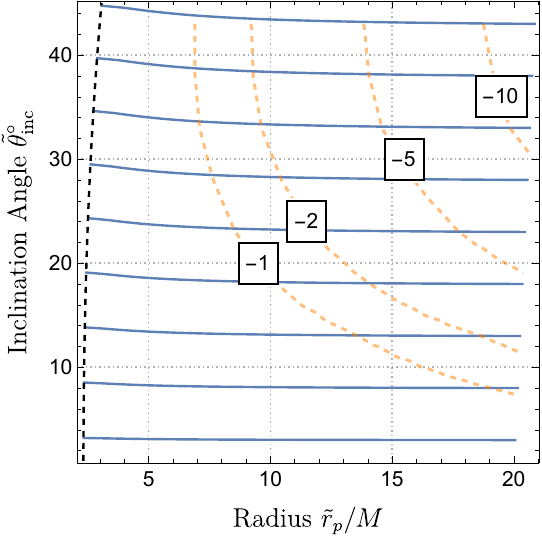}
	\caption{The blue curves show the adiabatic trajectories through $\{\r, \theta_\text{inc} \}$ space, while the orange contours denote the final value of $\varphi_\phi ^{(1)}$ when evolved from that point in the parameter space to the ISSO.}
	\label{fig:ParamterSpace}
\end{figure}

Since this is the first time that the full first-order self-force has been computed for quasi-spherical Kerr inspirals, we examine the impact that the adiabatic and post-adiabatic contributions have on the inspiral in Fig.~\ref{fig:ParamterSpace}. 
The blue curves show typical adiabatic trajectories through $\{\r,\theta_\text{inc}\}$ space. 
From this we see that the self-force causes the orbital radius to decrease over time, but also cause the inclination angle to increase slightly throughout the inspiral.
This is consistent with previous work on adiabatic quasi-spherical inspirals~\cite{Ryan:1995xi,Hughes2001}.
The post-adiabatic contributions to the solutions for $\r$ and $\theta_\text{inc}$ do not change this trend in any significant way since their contribution to the orbital elements is $\mathcal{O}(\mr)$.
However, the post-adiabatic contribution to the orbital phases is of order $\mathcal{O}(\mr^0)$ and is thus very significant.
This is demonstrated by the dashed contours on Fig.~\ref{fig:ParamterSpace}, which indicate the final value of the post-adiabatic piece of the azimuthal phase $\varphi_\phi^{(1)}$, when evolved from a given point in the parameter space to the ISSO.
Since it is the azimuthal phase that most strongly impacts the final waveform phase, this gives a good estimate of how many radians one can expect the waveform to be out of phase in the absence of post-adiabatic contributions.
It is worth restating that this does not include the second-order GSF contributions to $\varphi_\phi^{(1)}$ as these are not currently known.

\section{Discussion} \label{section:Discussion}

This paper presents the first calculations of the first-order gravitational self-force in the radiation gauge for spherical orbits in Kerr spacetime by utilizing a modified version of the code found in Refs.~\cite{vandeMeent:2015lxa,vandeMeent:2016pee,vandeMeent:2017bcc}. The main new element in this calculation is the inclusion of the gauge completion piece that puts the result in the correct asymptotically flat gauge.

To produce a continuous model this data is interpolated using Fourier decomposition and Chebyshev interpolation. 
Using only 162 points we obtain a model  with sub-percent accuracy for inclinations up to $0^\circ \leq \theta_\text{inc} \leq 45^\circ$.
This same method also allows interpolation of the orbit averaged rate of change of energy and angular momentum from the asymptotic fluxes for all inclinations (both prograde and retrograde) with an accuracy of $\sim 10^{-6}$ using only 342 points.
This is a significant improvement over other interpolation methods found in the literature which require at least an order of magnitude more points to achieve a comparable level of accuracy~\cite{Warburton2012,Osburn2016}.
This could be further improved with a better choice when rescaling the data before interpolation, ideally choosing a function informed by the leading-order PN contributions in the weak fields and/or the analytic structure of the GSF near the ISSO.
So far this model is only valid for orbits where the primary's spin is $a = 0.9 M$ and so interpolating over the other values of spin is left for future work. 
However this work, along with paper II and Ref.~\cite{Skoupy:2022adh}, shows that the Chebyshev interpolation methods are a promising approach for interpolating information from expensive GSF and flux codes across the vast four-dimensional parameter space of generic Kerr inspirals. 

Using our interpolated GSF model with the osculating geodesics (OG) formulation outlined in Ref.~\cite{Gair2011} and paper II along with the spherical limit derived in Appendix~\ref{section:OG_Spherical}, allows us to calculate the first ever quasi-spherical self-forced inspirals around a Kerr black hole.
Our Mathematica implementation of this will be made publicly available on the Black Hole Perturbation Toolkit~\cite{BHPToolkit}.
However, for a binary with a small mass ratio $\epsilon$, numerically evolving these inspirals can take minutes to hours due to the need to resolve the $\sim 1/\mr$ orbital oscillations. 

We overcome this by employing the technique of near-identity averaging transformations (NITs), as outlined in papers I and II, which produce equations of motion that capture the correct long-term secular evolution of the binary but can also be rapidly solved numerically. 
Following the methodology of Ref.~\cite{Pound2021}, we improve upon this formulation by employing a second averaging transformation such that the solutions to our equations of motion are parametrized by Boyer-Linquist coordinate time instead of Mino time.
Since Boyer-linquist time can be related to the time at the detector, this improved NIT procedure is much more convenient for generating waveforms and for data analysis. 
We also employ a two-timescale expansion (TTE) of the NIT equations of motion which factors out the dependence of the mass ratio, at the cost of doubling the number of equations to be solved.

The quantities calculated from either the NIT or TTE equations of motion remain close to the OG evolution variables throughout the inspiral to the expected order in the mass ratio, scaling similarly as the error expected from omitted higher order post-adiabatic corrections.
These averaging schemes work particularly well for quasi-spherical inspirals as we find that the mismatch between waveforms calculated from NIT or TTE inspirals and waveforms calculated from OG inspirals is $\leq 10^{-3}$, even  for binaries with $\mr \sim 10^{-1}$.

With our efficient inspiral trajectories, we investigate the effect of first-order self-force on inspirals across the spherical Kerr parameter space. 
We find that the orbital separation decreases with time while the orbit becomes more inclined over time, which is consistent with the findings of adiabatic evolutions of quasi-spherical Kerr inspirals. 
We also find that neglecting the conservative effects can result in an $\mathcal{O}(\mr^0)$ radian dephasing in the azimuthal phase, with this effect becoming larger with the inclination angle of the orbit.

We note that both NIT and TTE equations of motion allow us to further improve the accuracy of our waveforms by replacing the adiabatic terms with higher accuracy interpolants calculated from the asymptotic energy and angular momentum fluxes. 
Making this improvement can result in phase differences ranging from tens to tens of thousands of radians for multi-year-long EMRIs.
This highlights the necessity of efficient gravitational wave flux codes for both adiabatic and post-adiabatic EMRI waveforms~\cite{Hughes2021}.

This work only incorporates first-order GSF results.
Post-adiabatic waveforms will require second-order results not only to reach $\mathcal{O}(\mr^0)$ accuracy in the phases~\cite{Hinderer2008} but also, as was noted in paper II, to produce gauge-invariant waveforms~\cite{Lynch:2021ogr}. 
As such, the inspirals and waveforms produced here are only representative of the outgoing radiation gauge.

Thus, when second order effects are available for Kerr orbits, they should be folded into our framework. However, care should be taken as current second order calculations are produced assuming two-timescale expanded equations of motion and not assuming osculating geodesics and then performing the two-timescale expansion, as we have done.

We are also missing any post-adiabatic contributions due to the spin of the secondary~\cite{Skoupy:2021asz,Piovano:2020zin,Piovano:2021iwv,Mathews:2021rod,Drummond:2022xej,Drummond:2022efc,Skoupy:2023lih,Drummond:2023loz,Drummond:2023wqc}. 
Incorporating the linear-in-spin contribution to the energy and angular momentum fluxes into our averaged equations should be straightforward.
However, incorporating the conservative effects from the Mathisson-Papapetrou-Dixon equations will require careful consideration and will be the subject of future work.

We plan to extend this framework to the case of generic Kerr inspirals, but there are two challenges standing in our way. 
The first is the computational cost of generic Kerr GSF code coupled with the four-dimensional parameter space makes interpolating a GSF model impractical for now. 
The second challenge is the presence of transient self-force resonances where our NIT equations are formally singular which will force us to apply an alternate averaging procedure in their vicinity~\cite{Lukes-Gerakopoulos2021}.
There has been a lot of work covering the effects of transient resonances on EMRI trajectories due to the self-force~\cite{Flanagan2012a, Flanagan2012b,vandeMeent:2013sza, Berry2016,Mihaylov:2017qwn,Speri21,Nasipak:2021qfu} or an external third body~\cite{Gupta:2021cno,Gupta:2022fbe}.
However, evolving through self-force resonances while incorporating all self-force effects and retaining the subradian accuracy requirement remains an open problem.

Finally, we note we have used the semi-relativistic quadrupole formula to generate the waveforms from the OG, NIT and TTE inspirals. 
This is sufficient for this work as we only wish to compare the difference in the waveforms caused by different inspiral calculations.
However LISA data analysis will require fully relativistic waveform amplitudes such as those currently in the FastEMRIWaveforms (FEW) package for Schwarzschild inspirals~\cite{Chua2021a}.
Currently FEW only uses adiabatic inspirals, but this can be improved by employing either our NIT or TTE equations of motion.
Once the waveform amplitudes have been interpolated for Kerr inspirals, they can be combined immediately with the implementation presented in this work.

\begin{acknowledgments}
We thank Scott Hughes for sharing gravitational flux data for spherical orbits with us, and Lisa Drummond for a careful review of our osculating orbit evolution code.
We also thank Ian Hinder and Barry Wardell for the SimulationTools analysis package.
This work makes use of the Black Hole Perturbation Toolkit.
 PL acknowledges support from the Irish Research Council under Grant No.~GOIPG/2018/1978. 
 NW acknowledges support from a Royal Society–Science Foundation Ireland Research Fellowship.
 MvdM acknowledges financial support by the VILLUM Foundation (Grant No.~VIL37766) and the DNRF Chair program (Grant No.~DNRF162) by the Danish National Research Foundation.
 This publication has emanated from research conducted with the financial support of Science Foundation Ireland under Grant numbers 16/RS-URF/3428 and 17/RS-URF-RG/3490.
 For the purpose of open access, the author has applied a CC BY public copyright licence to any Author Accepted Manuscript version arising from this submission.
 The numerical calculation of the spherical orbit GSF was obtained using the IRIDIS High Performance Computing Facility at the University of Southampton.
\end{acknowledgments}


% \bibliographystyle{iop.bst}
\bibliography{SphericalKerrNITsReferences}

\appendix
%\begin{widetext}
\section{Osculating geodesic equations for spherical orbits} \label{section:OG_Spherical}

In this appendix we derive the evolution equation for the orbital radius $\r$ in the spherical limit due to the presence of a perturbing acceleration $\{a_t, a_r, a_z, a_{\phi} \}$. Using the chain rule, we obtain
\begin{equation} \label{eq:Chain_Rule}
	\frac{d \r}{d \lambda} = \PD{\r}{\mathcal{J}_j}\frac{d \mathcal{J}_j}{d \lambda}.
\end{equation}
Recalling the expressions for the evolution equations for the integrals of motion $ \vec{\mathcal{J}} = \{ \En, \Lz, \K \}$ given in Appendix B of paper II,
\begin{subequations}
	\begin{gather}
		\frac{d \En}{d\lambda} = - \pos{\Sigma} a_t, \quad
		\frac{d \Lz}{d\lambda} = \pos{\Sigma} a_{\phi},
		\tag{\theequation a-b}
	\end{gather}
	\begin{gather}
		\frac{d \K}{d \lambda} =  -\frac{2 \pos{\Sigma}}{\pos{\Delta}} \left( \pos{\varpi}^2 \pos{\mathcal{B}} a_t +  a \pos{\mathcal{B}}  a_\phi + \pos{\Delta}^2 u_r a_r \right).
		\tag{\theequation c}
	\end{gather}
\end{subequations}
To find the partial derivatives $\partial \r / \partial \mathcal{J}_j $, we make use of the value of the radial potential $V_r$ defined as
\begin{equation}
\begin{split}
	V_r(r) &= \mathcal{B}^2 - \Delta\left(r^2+ \K \right)  \\
		& = -(1-\En^2)(r - r_1) (r-r_2) (r-r_3)(r-r_4) \\
		&= -(1-\En^2)(r - \r)^2 (r-r_3)(r-r_4).
\end{split}
\end{equation}
where $r_1,r_2,r_3,r_4$ are the roots of $V_r$ and in the spherical case: $r_1 = r_2 = \r$.
We then define the derivative of $V_r$ with respect to $r$ as
\begin{equation} \label{eq:kappa_def}
	\begin{split}
	\kappa \coloneqq & \frac{d V_r}{d r} \\
		   = & 4 \En \mathcal{B}  r - 2 \Delta r - 2(r-M)(r^2 + K)\\
		   = & -2(1-\En^2)\bigg[ (r-r_3)(r-r_4) \\
		  	 &+ (r-\r)(r-r_4) + (r-\r)(r-r_3)\bigg].
	\end{split}
\end{equation}
If we take the derivative of $\kappa$ with respect to $r$ and evaluate at $r = \r$, we obtain
\begin{equation}
	\frac{d \pos{\kappa}}{d r} =  - 2 (1-\En^2)(\r-r_3)(\r-r_4).
\end{equation}
However, if we now take the partial derivatives of $\kappa$ with respect to $\vec{\mathcal{J}}$ and evaluate at $r = \r$ we are left with
\begin{equation}
	\begin{split}
	\PD{\pos{\kappa}}{\mathcal{J}_j}  &=  2 (1-\En^2)(\r-r_3)(\r -r_4) \PD{\r}{\mathcal{J}_j} \\
	&= -\frac{d \pos{\kappa}}{d r}  \PD{\r}{\mathcal{J}_j}.
	\end{split}
\end{equation}
Substituting the above result into Eq.~\eqref{eq:Chain_Rule}, explicitly calculating $\partial \kappa / \partial \mathcal{J}_j$ using Eq.~\eqref{eq:kappa_def}, evaluating at $r = \r$ and simplifying gives us our final result
\begin{align}
	\begin{split}
		\frac{d \r}{d \lambda }
		 & =   -4 \r \frac{
		 	(\pos{\mathcal{B}} + \pos{\varpi}^2 \En) \frac{d E}{d \lambda} - a  \En \frac{d \Lz}{d \lambda} + \frac{M - \r}{2\r} \frac{d \K}{ d \lambda}
	 	}{
	 	\frac{d \pos{\kappa}}{d r}
 		}
		\\& \coloneqq F_r^{(1)}.
	\end{split}
\end{align}
The above relationship is finite for all bound stable spherical orbits and only becomes ill-defined when $d\pos{\kappa}/ d r = 0$, which, only occurs when $r_p = r_3$ , i.e., when $ \r = r_{\text{ISSO}}$.

\section{Equivalence between $\vec{\varphi}$ and waveform voices} \label{section:tNIT_and_Wavefrom_Phases}
In this appendix, we wish to establish a relationship between the solutions for the inspiral quantities and the waveform.
We start by assigning the origin of our coordinate scheme to be the position of the primary MBH. 
We then set the Cartesian coordinates of our observer to be $x = (t_\text{ret},\vec{x})$. 
It is useful to express the observer's coordinates in terms of spherical coordinates $(R, \theta, \Phi)$, where $R$ is the distance from the observer to the origin given by $R^2 = \vec{x} \cdot \vec{x}$, $\Theta$ is the observer's latitude and $\Phi$ is the observer's azimuth.
We assign the Cartesian coordinates of the secondary, which we model as a point particle, to be $x_p = (t_p, \vec{x}_p)$.
As such, we can express the retarded time as measured by the observer in terms of Boyer-Lindquist coordinate time $t$ via $t_\text{ret} = t - |\vec{x} - \vec{x}_p| \approx t - R $ in the limit where $R$ is large.
The complex waveform strain can be decomposed onto a basis of spin-weighted spherical harmonics $_{-2} Y_{lm}$ given by: 

\begin{equation}
	h(t_\text{ret}) = h_+ - i h_\times = \frac{1}{R} \sum_{lm} H_{lm}(t_\text{ret}) _{-2} Y_{lm} (\Theta, \Phi).
\end{equation}
If we assume the secondary is moving on a geodesic, the waveform modes $H_{lm}$ exhibit a discrete frequency spectrum and so can be Fourier decomposed into:
\begin{equation}
	H_{lm}(t_\text{ret}) =  \sum_{(n,k)}  A_{l m n k} (\vec{P}) e^{ -i (m  \Omega_{\phi}^{(0)} + n  \Omega_{r}^{(0)} + k  \Omega_{z}^{(0)}  ) t_\text{ret}}
\end{equation}
where the complex amplitudes $A_{l m n k}$ can be related to non-homogeneous Teukolsky amplitudes, which can be precomputed for a given set of orbital elements $\vec{P}$ ~\cite{Hughes2021}.
Such waveforms are known as ``snapshot" waveforms, as they only capture a small section of the total waveform~\cite{Drasco:2005kz}. 

For a full EMRI waveform, one needs to account for the fact that the frequencies and the orbital elements will slowly evolve with time, resulting in a continuous frequency spectrum. 
As such, this Fourier mode decomposition becomes a ``multi-voice" decomposition~\cite{Hughes2021}:
\begin{equation}
	H_{lm}(t_\text{ret}) =  \sum_{(n,k)}  A_{l m n k}(\vec{P}(t_\text{ret})) e^{ -i \varPhi_{m n k }(t_\text{ret})},
\end{equation}
where the waveform ``voices" $\varPhi_{m n k}$ are given by 
\begin{equation}
	\varPhi_{m n k} = m \int_0^t \Omega_\phi(t) dt + n \int_0^t \Omega_r(t) dt + k \int_0^t \Omega_z(t) dt.
\end{equation}
Let us recall the equations of motion for the orbital phases obtained after performing the NIT and the additional transformation such that our solutions are in terms of Boyer-Lindquist time \eqref{eq:t_transformed_EoM}. 
When we express this in integral form, one obtains
\begin{align}
	\begin{split}
	\varphi_\alpha & = \int_0^t \left(\Omega^{(0)}_\alpha(t) +  \mr \Omega^{(1)}_\alpha(t) + \mathcal{O}(\mr^2) \right) dt 
	\\ & = \int_0^t \Omega_\alpha(t) dt + \mathcal{O}(\mr)
\end{split}
\end{align}
As such we can see that to leading order in $\mr$, these phases are directly related to the waveform voices:
\begin{equation}
	\varPhi_{m n k} = m \varphi_\phi + n \varphi_r + k \varphi_z + \mathcal{O}(\mr)
\end{equation}
This is further supported by Ref.~\cite{McCart2021}, where a relationship between the waveform phases and the NIT phases $\vec{\nit{q}}$ was independently derived.
When expressed in our notation with polar motion included, this relationship is given by:
\begin{equation}
\begin{split}
		\varPhi_{m n k} &= m \nit{\phi} + n \nit{q}_r + k \nit{q}_z\\
		&\quad +  (m \Omega_\phi^{(0)} + n \Omega_r^{(0)} + k \Omega_z^{(0)})(t - \nit{t}) + \mathcal{O}(\mr)
\end{split}
\end{equation}
Using the relationship between $t$ and $\nit{t}$ given by Eq.~\eqref{eq:Extrinsic_Transformation}, the result for $\Delta t$ given by Eq.~\eqref{eq:Delta_t}, and the relationship between the NIT action angles given by Eq.~\eqref{eq:Phase_Transformation}, one can obtain to leading order in $\mr$,
\begin{widetext}
\begin{equation}
	\begin{split}
		\varPhi_{m n k } &= m (\nit{\phi} - \Omega^{(0)}_\phi (t - \nit{t})) + n (\nit{q}_r - \Omega^{(0)}_r (t - \nit{t})) + k (\nit{q}_z - \Omega_z (t - \nit{t})) +\mathcal{O}(\mr)\\
		& = m (\nit{\phi} - \Omega^{(0)}_\phi Z_t^{(0)}) + n (\nit{q}_r - \Omega^{(0)}_r Z_{t}^{(0)}) + k (\nit{q}_z - \Omega^{(0)}_z Z_{t}^{(0)} +\mathcal{O}(\mr)) \\
		&  = m (\nit{\phi} + \Omega^{(0)}_\phi \Delta t) + n (\nit{q}_r + \Omega^{(0)}_r \Delta t) + k (\nit{q}_z + \Omega^{(0)}_z \Delta t) +\mathcal{O}(\mr)\\
		& =  m (\nit{\phi} + \Delta \varphi_\phi) + n (\nit{q}_r + \Delta \varphi_r) + k (\nit{q}_z + \Delta \varphi_z) +\mathcal{O}(\mr) \\
		& = m \varphi_{\phi} + n \varphi_r + k \varphi_z +\mathcal{O}(\mr)
	\end{split}
\end{equation}
One can also freely replace the dependence of the evolving orbital elements $\vec{P}(t)$ with either $\vec{\nit{P}}(t)$ or $\vec{\mathcal{P}}(t)$, as at leading order in $\mr$ they are identical.
As a result, we can now relate a solution to the EMRI's inspiral trajectory to its associated Teukolsky based waveform.

\section{Gauge completion integrands}\label{app:gaugecomp}

Here we give the explicit integrands for the gauge completion integrals~\eqref{eq:gaugecompinta}~and~\eqref{eq:gaugecompintb},
\begin{equation}
\mathcal{I}_\alpha = \begin{aligned}[t]
&\mathcal{E}^2 \r \left(1-\z^2\right) 
\Big[
	3 \r^{10}
	+3 \r^9 \left(1-\z^2\right) 
	-6 \r^8  a^2 \left(1 + \z^2 - 3 \z^4 \right)
	-2\r^7  a^2 \left(5+4 \z^2+3 \z^4\right)
	\\
	&\qquad-\r^6 a^2 \left(3 a^2 -(48-31a^2)\z^2 + (24-47 a^2)\z^4  - 3 a^2 \z^6\right) 
	-\r^5  a^4 \left(5 - 9 \z^2 + 41 \z^4 + 3 \z^6\right)
	\\
	&\qquad-2 \r^4  a^4 \z^2 \left(2(1+2 a^2) - (4+21a^2) \z^2 -(a^2 +12) \z^4\right)
	+2 \r^3 a^6 \z^2\left(11-16 \z^2+\z^4\right) 
	\\
	&\qquad-\r^2 a^6 \z^2 \left(10 a^2-15 a^2 \z^2+5 a^2 \z^4+8 \z^4\right) 
	-\r a^8 \z^4 \left(5-13 \z^2\right) 
	-2 a^{10} \z^4 \left(1-2 \z^2\right) \left(1-\z^2\right) 
\Big]
\\
&+(\Lz-a \mathcal{E})^2 \left(1-\z^2\right)^2 
 \Big[
 	3 \r^8
	-3 \r^7 \left(4-a^2 +5 a^2 \z^2\right)
 	-6 \r^6 a^2 \left(3+\z^2\right)
	\\
	&\qquad+\r^5 a^2 \left(4 +6 a^2+ 3(12- 13 a^2 ) \z^2 - 3 a^2 \z^4\right)
	-\r^4 a^4 \left(13 - 12 \z^2 + 9 \z^4\right)
	\\
	&\qquad+3 \r^3 a^4 \left(a^2 - (4+11 a^2) \z^2 - 2 a^2 \z^4\right)
 	+2 \r^2 a^6\z^2 \left(9- \z^2\right)
 	-3 \r a^8 \z^2 \left(3+\z^2\right)
 	-a^8 \z^4
 \Big]
\\
&-\Lz^2 \r 
\Big[
	3\r^8 \left(1+\z^2\right)^2
	-3 \r^7 \left(1+3 \z^2\right) \left(3+\z^2\right)
	-\r^6 \left(a^2  - (48-2 a^2)\z^2 - 47 a^2 \z^4 +12 a^2 \z^6  \right)
	\\
	&\qquad-\r^5 a^2 \left(5 -8 \z^2 + 65 \z^4 + 18 \z^6\right)
	-\r^4 a^2 \z^2 \left(16 +26 a^2 - 57 a^2 \z^2 - (48-4 a^2) \z^4 + 3 a^2 \z^6\right)
	\\
	&\qquad+\r^3 a^4 \z^2 \left(22 -21 \z^2 -8 \z^4 - 9 \z^6\right)
	-\r^2 a^4 \z^2 \left(10 a^2  - 15 a^2 \z^2  + 8(2 + a^2) \z^4 - 3 a^2 \z^6\right)
	\\
	&\qquad-\r a^6 \z^4 \left(1-5 \z^2\right) \left(5-\z^2\right)
	+2 a^8 \z^4 \left(1 - 4 \z^2 + \z^4\right)
\Big]
\\
&-\r \left(3 \r^2-a^2\right) \left(1-\z^4\right) \left(a^2 \z^2-\mathcal{Q}\right) \pos{\Delta}^2 \pos{\Sigma},
\end{aligned}
\end{equation}
and
\begin{equation}
\mathcal{I}_\beta = \begin{aligned}[t]
&\mathcal{E}^2 a^2 \r \left(1-\z^2\right) 
\Big[
	\r^9 \left(1-\z^2\right)  \left(1-9 \z^2\right) 
	-\r^8 \left(6  - (18 + a^2)\z^2  + (12- a^2) \z^4\right)
	\\
	&\qquad
	-\r^7 a^2 \left(2 + 35 \z^2 - 32 \z^4 + 3 \z^6 \right)
	-\r^6 a^2 \left(6 - (46+a^2) \z^2 + 2(22-a^2)\z^4 - (12+a^2)\z^6 \right)
	\\ &\qquad
	+\r^5 a^4 \left(1 - 28 \z^2 + 33 \z^4 - 14 \z^6\right)
	-\r^4 a^4  \z^2 \left(1 - \z^2\right) \left(4+a^2 +(26+a^2) \z^2 \right)
	\\ &\qquad
	+\r^3 a^6 \z^2 \left(1 + 18 \z^2 - 11 \z^4\right)
	-\r^2 a^6 \z^2 \left(a^2 + 2(3+a^2) \z^2 + (2+a^2)\z^4\right)
	+(4 \r -a^{2} ) a^8 \z^4 \left(1+\z^2\right)
\Big] 
\\
&+(\Lz-a \mathcal{E})^2 \left(1-\z^2\right)^2 
\Big[
	3(\r-2)\r^9
	+3 \r^8 a^2 \left(4-3 \z^2\right)
	-2 \r^7 a^2 \left(11-6 \z^2\right)
	+\r^6 a^4 \left(7-33\z^2\right)
	\\
	&\qquad-2 \r^5 a^4 \left(4 - 18 \z^2 + 3 \z^4\right)
	+\r^4  a^6\left(2 - 27 \z^2 - \z^4\right)
	-6 \r^3 a^6 \z^4
	-\r^2 a^8 \z^2 \left(3-2 \z^2\right)
	-\z^4 a^{10}
\Big]
\\
&-\Lz^2 \r 
\Big[
	3 \r^9 \left(1-\z^2\right)^2
	-\r^8 \left(6-a^2 -2(6+a^2)\z^2+(6-a^2) \z^4\right)
	+\r^7  a^2 \left(2 - 29 \z^2 + 20 \z^4 - 9 \z^6\right)
	\\
	&\qquad-\r^6 a^2 \left(6+a^2 -(40+3 a^2)\z^2 + 5(6-a^2) \z^4 - (12+a^2) \z^6\right)
	+\r^5 a^4 \left(1 - 14\z^2 + 25 \z^4 - 28 \z^6\right)
	\\
	&\qquad	-2 \r^4 a^4 \z^2  \left(1-\z^2\right) \left(2+2 a^2 +(17+2 a^2)\z^2 - 3\z^4\right)
	+\r^3 a^6 \z^2\left(2 \z^6-19 \z^4+32 \z^2+1\right)
	\\
	&\qquad	-\r^2 a^6 \z^2 \left(a^2 + (6+5a^2)\z^2 + (4+3 a^2)\z^4 + (6-a^2) \z^6\right)
	+(4 \r- a^{2}) a^8 \z^4 \left(1+\z^2\right)^2
\Big]
\\
&-a^2  \r \left(\r^2-a^2\right)   \left(1 - \z^4\right)  (a^2 \z^2-\mathcal{Q})\pos{\Delta}^2 \pos{\Sigma}.
\end{aligned}
\end{equation}

\end{widetext}
\end{document}